

%
\newbox\leftpage \newdimen\fullhsize \newdimen\hstitle \newdimen\hsbody
\tolerance=1000\hfuzz=2pt
\def\bigans{b }
\def\answ{b } 
\ifx\answ\bigans\message{(This will come out unreduced.}
\magnification=1200\baselineskip=18pt plus 2pt minus 1pt
\hsbody=\hsize \hstitle=\hsize 
\else\def\apans{l }\message{ lyman or hepl (l/h) (lowercase!) ? }
\read-1 to \apansw\message{(This will be reduced.}
\let\lr=L
\magnification=1000\baselineskip=16pt plus 2pt minus 1pt
\voffset=-.31truein\vsize=7truein
\hstitle=8truein\hsbody=4.75truein\fullhsize=10truein\hsize=\hsbody
\ifx\apansw\apans\special{ps: landscape}\hoffset=-.59truein
  \else\hoffset=.05truein\fi
\output={\ifnum\pageno=0 
  \shipout\vbox{\hbox to \fullhsize{\hfill\pagebody\hfill}}\advancepageno
  \else
  \almostshipout{\leftline{\vbox{\pagebody\makefootline}}}\advancepageno
  \fi}
\def\almostshipout#1{\if L\lr \count1=1
      \global\setbox\leftpage=#1 \global\let\lr=R
  \else \count1=2
    \shipout\vbox{\ifx\apansw\apans\special{ps: landscape}\fi 
      \hbox to\fullhsize{\box\leftpage\hfil#1}}  \global\let\lr=L\fi}
\fi
%
\catcode`\@=11 
\newcount\yearltd\yearltd=\year\advance\yearltd by -1900

\def\Title#1#2{\nopagenumbers\abstractfont\hsize=\hstitle\rightline{#1}%
\vskip 1in\centerline{\titlefont #2}\abstractfont\vskip .5in\pageno=0}
\def\Date#1{\vfill\leftline{#1}\tenpoint\supereject\global\hsize=\hsbody%
\footline={\hss\tenrm\folio\hss}}
\def\draftmode{\def\draftdate{{\rm preliminary draft:
\number\month/\number\day/\number\yearltd\ \ \hourmin}}%
\headline={\hfil\draftdate}\writelabels\baselineskip=20pt plus 2pt minus 2pt
{\count255=\time\divide\count255 by 60 \xdef\hourmin{\number\count255}
	\multiply\count255 by-60\advance\count255 by\time
   \xdef\hourmin{\hourmin:\ifnum\count255<10 0\fi\the\count255}}}

\def\nolabels{\def\eqnlabel##1{}\def\eqlabel##1{}\def\reflabel##1{}}
\def\writelabels{\def\eqnlabel##1{%
{\escapechar=` \hfill\rlap{\hskip.09in\string##1}}}%
\def\eqlabel##1{{\escapechar=` \rlap{\hskip.09in\string##1}}}%
\def\reflabel##1{\noexpand\llap{\string\string\string##1\hskip.31in}}}
\nolabels
%
\global\newcount\secno \global\secno=0
\global\newcount\meqno \global\meqno=1
\def\newsec#1{\global\advance\secno by1
\xdef\secsym{\the\secno.}\global\meqno=1
\bigbreak\bigskip
\noindent{\bf\the\secno. #1}\par\nobreak\medskip\nobreak}
\xdef\secsym{}
%
\def\appendix#1#2{\global\meqno=1\xdef\secsym{\hbox{#1.}}\bigbreak\bigskip
\noindent{\bf Appendix #1. #2}\par\nobreak\medskip\nobreak}
%
%
\def\eqnn#1{\xdef #1{(\secsym\the\meqno)}%
\global\advance\meqno by1\eqnlabel#1}
\def\eqna#1{\xdef #1##1{\hbox{$(\secsym\the\meqno##1)$}}%
\global\advance\meqno by1\eqnlabel{#1$\{\}$}}
\def\eqn#1#2{\xdef #1{(\secsym\the\meqno)}\global\advance\meqno by1%
$$#2\eqno#1\eqlabel#1$$}
%
\newskip\footskip\footskip14pt plus 1pt minus 1pt 
\def\f@@t{\baselineskip\footskip\bgroup\aftergroup\@foot\let\next}
\setbox\strutbox=\hbox{\vrule height9.5pt depth4.5pt width0pt}
\global\newcount\ftno \global\ftno=0
\def\foot{\global\advance\ftno by1\footnote{$^{\the\ftno}$}}
%
%
\global\newcount\refno \global\refno=1
\newwrite\rfile
\def\ref{[\the\refno]\nref}
\def\nref#1{\xdef#1{[\the\refno]}\ifnum\refno=1\immediate
\openout\rfile=refs.tmp\fi\global\advance\refno by1\chardef\wfile=\rfile
\immediate\write\rfile{\noexpand\item{#1\ }\reflabel{#1}\pctsign}\findarg}
\def\findarg#1#{\begingroup\obeylines\newlinechar=`\^^M\pass@rg}
{\obeylines\gdef\pass@rg#1{\writ@line\relax #1^^M\hbox{}^^M}%
\gdef\writ@line#1^^M{\expandafter\toks0\expandafter{\striprel@x #1}%
\edef\next{\the\toks0}\ifx\next\em@rk\let\next=\endgroup\else\ifx\next\empty%
\else\immediate\write\wfile{\the\toks0}\fi\let\next=\writ@line\fi\next\relax}}
\def\striprel@x#1{} \def\em@rk{\hbox{}} {\catcode`\%=12\xdef\pctsign{
\def\semi{;\hfil\break}
\def\addref#1{\immediate\write\rfile{\noexpand\item{}#1}} 
\def\listrefs{
\vfill\eject\immediate\closeout\rfile
\baselineskip=24pt\centerline{{\bf References}}\bigskip{\frenchspacing%
\escapechar=` \input refs.tmp\vfill\eject}\nonfrenchspacing}
\def\startrefs#1{\immediate\openout\rfile=refs.tmp\refno=#1}
\def\figures{\centerline{{\bf Figure Captions}}\medskip\parindent=40pt}
\def\fig#1#2{\medskip\item{Figure ~#1:  }#2}
\catcode`\@=12 
%
\ifx\answ\bigans
\font\titlerm=cmr10 scaled\magstep3 \font\titlerms=cmr7 scaled\magstep3
\font\titlermss=cmr5 scaled\magstep3 \font\titlei=cmmi10 scaled\magstep3
\font\titleis=cmmi7 scaled\magstep3 \font\titleiss=cmmi5 scaled\magstep3
\font\titlesy=cmsy10 scaled\magstep3 \font\titlesys=cmsy7 scaled\magstep3
\font\titlesyss=cmsy5 scaled\magstep3 \font\titleit=cmti10 scaled\magstep3
\else
\font\titlerm=cmr10 scaled\magstep4 \font\titlerms=cmr7 scaled\magstep4
\font\titlermss=cmr5 scaled\magstep4 \font\titlei=cmmi10 scaled\magstep4
\font\titleis=cmmi7 scaled\magstep4 \font\titleiss=cmmi5 scaled\magstep4
\font\titlesy=cmsy10 scaled\magstep4 \font\titlesys=cmsy7 scaled\magstep4
\font\titlesyss=cmsy5 scaled\magstep4 \font\titleit=cmti10 scaled\magstep4
\font\absrm=cmr10 scaled\magstep1 \font\absrms=cmr7 scaled\magstep1
\font\absrmss=cmr5 scaled\magstep1 \font\absi=cmmi10 scaled\magstep1
\font\absis=cmmi7 scaled\magstep1 \font\absiss=cmmi5 scaled\magstep1
\font\abssy=cmsy10 scaled\magstep1 \font\abssys=cmsy7 scaled\magstep1
\font\abssyss=cmsy5 scaled\magstep1 \font\absbf=cmbx10 scaled\magstep1
\skewchar\absi='177 \skewchar\absis='177 \skewchar\absiss='177
\skewchar\abssy='60 \skewchar\abssys='60 \skewchar\abssyss='60
\fi
\skewchar\titlei='177 \skewchar\titleis='177 \skewchar\titleiss='177
\skewchar\titlesy='60 \skewchar\titlesys='60 \skewchar\titlesyss='60
\def\titlefont{\def\rm{\fam0\titlerm}
\textfont0=\titlerm \scriptfont0=\titlerms \scriptscriptfont0=\titlermss
\textfont1=\titlei \scriptfont1=\titleis \scriptscriptfont1=\titleiss
\textfont2=\titlesy \scriptfont2=\titlesys \scriptscriptfont2=\titlesyss
\textfont\itfam=\titleit \def\it{\fam\itfam\titleit} \rm}
\ifx\answ\bigans\def\abstractfont{\tenpoint}\else
\def\abstractfont{\def\rm{\fam0\absrm}
\textfont0=\absrm \scriptfont0=\absrms \scriptscriptfont0=\absrmss
\textfont1=\absi \scriptfont1=\absis \scriptscriptfont1=\absiss
\textfont2=\abssy \scriptfont2=\abssys \scriptscriptfont2=\abssyss
\textfont\itfam=\bigit \def\it{\fam\itfam\bigit}
\textfont\bffam=\absbf \def\bf{\fam\bffam\absbf} \rm} \fi
\def\tenpoint{\def\rm{\fam0\tenrm}
\textfont0=\tenrm \scriptfont0=\sevenrm \scriptscriptfont0=\fiverm
\textfont1=\teni  \scriptfont1=\seveni  \scriptscriptfont1=\fivei
\textfont2=\tensy \scriptfont2=\sevensy \scriptscriptfont2=\fivesy
\textfont\itfam=\tenit \def\it{\fam\itfam\tenit}
\textfont\bffam=\tenbf \def\bf{\fam\bffam\tenbf} \rm}
%
%
\def\noblackbox{\overfullrule=0pt}
\hyphenation{anom-aly anom-alies coun-ter-term coun-ter-terms}
\def\inv{^{\raise.15ex\hbox{${\scriptscriptstyle -}$}\kern-.05em 1}}
\def\dup{^{\vphantom{1}}}
\def\Dsl{\,\raise.15ex\hbox{/}\mkern-13.5mu D} 
\def\dsl{\raise.15ex\hbox{/}\kern-.57em\partial}
\def\del{\partial}
\def\Psl{\dsl}
\def\tr{{\rm tr}} \def\Tr{{\rm Tr}}
\font\bigit=cmti10 scaled \magstep1
\def\biglie{\hbox{\bigit\$}} 
\def\lspace{\ifx\answ\bigans{}\else\qquad\fi}
\def\lbspace{\ifx\answ\bigans{}\else\hskip-.2in\fi} 
\def\boxeqn#1{\vcenter{\vbox{\hrule\hbox{\vrule\kern3pt\vbox{\kern3pt
	\hbox{${\displaystyle #1}$}\kern3pt}\kern3pt\vrule}\hrule}}}
\def\mbox#1#2{\vcenter{\hrule \hbox{\vrule height#2in
		\kern#1in \vrule} \hrule}}  
%
\def\CAG{{\cal A/\cal G}}   
\def\CA{{\cal A}} \def\CC{{\cal C}} \def\CF{{\cal F}} \def\CG{{\cal G}}
\def\CL{{\cal L}} \def\CH{{\cal H}} \def\CI{{\cal I}} \def\CU{{\cal U}}
\def\CB{{\cal B}} \def\CR{{\cal R}} \def\CD{{\cal D}} \def\CT{{\cal T}}
\def\e#1{{\rm e}^{^{\textstyle#1}}}
\def\grad#1{\,\nabla\!_{{#1}}\,}
\def\gradgrad#1#2{\,\nabla\!_{{#1}}\nabla\!_{{#2}}\,}
\def\ph{\varphi}
\def\psibar{\overline\psi}
\def\om#1#2{\omega^{#1}{}_{#2}}
\def\vev#1{\langle #1 \rangle}
\def\lform{\hbox{$\sqcup$}\llap{\hbox{$\sqcap$}}}
\def\darr#1{\raise1.5ex\hbox{$\leftrightarrow$}\mkern-16.5mu #1}
\def\lie{\hbox{\it\$}} 
\def\ha{{1\over2}}
\def\half{{\textstyle{1\over2}}} 
\def\roughly#1{\raise.3ex\hbox{$#1$\kern-.75em\lower1ex\hbox{$\sim$}}}

\Title{PUPT-1495, IN94020,  hep-ph/9410281}
{\vbox{\centerline{Non-local Electroweak Baryogenesis}
	\vskip2pt\centerline{Part I : Thin Wall Regime}}}

\baselineskip 18pt
\centerline{{\bf Michael Joyce}\footnote{$^\dagger$}{e-mail:
joyce@nxth01.cern.ch, tomislav@hepth.cornell.edu, \hfil\break
neil@puhep1.princeton.edu}}
\centerline{{\bf Tomislav Prokopec}{$^\dagger$}}
\centerline{and}
\centerline{{\bf Neil Turok}{$^\dagger$}}
\centerline{Joseph Henry Laboratories }
\centerline{Princeton University}
\centerline{Princeton, NJ 08544.}
\vskip .2 in
\centerline{\bf Abstract}
\baselineskip 12pt
We investigate `non-local' schemes for baryogenesis at a first order
electroweak phase transition, in which the effects of
a $CP$ violating condensate on the bubble wall propagate into
the unbroken phase where the sphaleron rate is unsupressed.
Such a condensate exists in multi-Higgs extensions of the standard
model, and may exist due to an instability in the minimal standard
model. In this paper we first discuss the general problem of
determining the $CP$ violating perturbations, distinguishing
two regimes (quantum and classical). We then give  an analytic
treatment of quantum mechanical reflection in the  thin wall
regime, in which interactions with the plasma can be neglected as
a particle propagates across the wall. We focus  on leptons
because of their much weaker coupling to the plasma. We argue that
they are likely to be  accurately described by this calculation,
but quarks are not. The relative magnitude of the baryon asymmetry
produced for different fermions depends on their relative Yukawa
couplings ({\it not} their zero temperature masses), their transport
properties and their interactions. We calculate the baryon asymmetry
for various parameter ranges and conclude that asymmetries comparable
with observations can be generated.

\Date{Revised: October 1995}

\baselineskip 12pt 

\medskip

\centerline { \bf 1. Introduction}
Almost twenty years passed before it was realized  that the conditions
stated by Sakharov
for the dynamical generation of the baryon asymmetry
in the universe were actually satisfied in the standard
Weinberg-Salam model
\ref\review{For reviews see N. Turok, in {\it Perspectives in Higgs Physics},
ed. G. Kane, publ. World Scientific, p. 300(1992);
 A. Cohen, D. Kaplan and A. Nelson, Ann. Rev. Nucl. Part. Phys.
{\bf 43} 27(1993).}.
The required baryon number violating processes arise from
the chiral nature of the theory
and the topology of the vacuum,
as encoded in the chiral $SU(2)_L$ anomaly
\ref\KRS{G. 't Hooft, Phys. Rev. Lett. {\bf 37}, 8 (1976);
V. Kuzmin, V. Rubakov and M. Shaposhnikov, Phys. Lett. {\bf 155B},
 36 (1985); F. Klinkhamer and N. Manton, Phys. Rev. {\bf D30}, 2212 (1984);
 P. Arnold and L. McLerran, Phys. Rev. {\bf D37}, 1020.}.
C and CP violation are built into the standard model,  and
Sakharov's final
ingredient, a departure from thermal equilibrium, is
believed to be provided by the first-order nature of the electroweak
phase transition.

Despite the fact that the minimal standard model
has all of Sakharov's ingredients,
it has proven difficult to use it
to produce the required asymmetry. The main obstacle to
its success is the very small CP violation in the KM matrix. A
second problem is ensuring that the baryon number violating processes
are suppressed after the phase transition - this
translates into a rather low upper bound on the
Higgs mass $M_H{<\over ~} 35-80$ GeV,
potentially in conflict with the experimental lower bound.
The exact constraint is still rather uncertain,
because it relies on the details of
the
finite temperature effective potential,
 which has only recently begun to be be  seriously studied
\ref\dhll{M. Dine, P. Huet,
R. Leigh and A. Linde, Phys. Lett. B283 (1992) 319;
Phys. Rev. D 46 (1992) 550.};
\ref\bunk{B. Bunk, E.-M. Ilgenfritz, J. Kripfganz and A. Schiller,
Phys. Lett. {\bf B284}, 371 (1992); Nuc. Phys. {\bf B403}, 453 (1993).},
\ref\dinebag{M. Dine and J. Bagnasco, Phys. Lett. B303 (1993) 308.},
\ref\arnold{P. Arnold and O. Espinosa, Phys. Rev. {\bf D47}, 3546
(1993).},
\ref\nonpert{K. Kajantie, K. Rummukainen and M. Shaposhnikov,
Nucl. Phys. {\bf B407}, 356 (1993);
F. Farakos, K. Kajantie, K. Rummukainen and M. Shaposhnikov,
Phys. Lett. {\bf B 336} 494 (1994), hep-ph/9405234;
Preprint CERN-TH.7220/94 (1995);
Z. Fodor, J. Hein, K. Jansen, A. Jaster, I. Montvay and F. Csikor
Phys. Lett. {\bf B 334}, 405 (1994), hep-lat/9405021; also hep-lat/9411052;
W. Buchm\"{u}ller and Z. Fodor, Phys. Lett.
{\bf B331}, 124 (1994).}. A further problem for baryogenesis in
in the minimal standard model is that for the observed top quark
mass, unless the Higgs is relatively heavy, $M_H{> \over ~} 130 $ GeV,
the {\it zero} temperature effective
potential is unbounded from below, and this leads to an unacceptable
instability \ref\sher{M. Sher, Phys. Lett.{\bf 317B},159 (1993)
ADDENDUM-ibid.B331:448,1994} .

One notable
recent attempt to elaborate a minimal standard model mechanism
is that of Farrar and Shaposhnikov
\ref\FarrarShaposhnikov{G. Farrar and M. Shaposhnikov,
Phys. Rev. Lett.{\bf  70}, 2833 (1993); Phys. Rev. {\bf D50}, 774 (1994).},
who make use of subtle finite temperature effects on the
reflection of quarks from the bubble walls formed at the phase transition.
This  calculation has been argued in
\ref\ghop{M. B. Gavela, P. Hernandez, J. Orloff and O. Pene,
Mod. Phys. Lett. {\bf 9A}, 795 (1994); M. B. Gavela, M. Losano,
J. Orloff and O. Pene, Nucl. Phys. {\bf B430}, 345 and 382 (1994),
hep-ph/9406288 and 9406289.} and
\ref\hs{ P. Huet and E. Sather,  Phys. Rev. {\bf D 51}, 379 (1995),
hep-ph/9404302.}
to greatly overestimate the baryon asymmetry, because it
neglects the
imaginary part of the quark self-energy.
A suggestion as to how
CP violation
could be amplified in the minimal standard model was made
by Nasser and Turok
\ref\snnt{S. Nasser and N. Turok, Princeton preprint PUPT-1456 (1994),
 hep-ph/9406270.}.
The idea here  is
that CP could be spontaneously broken by the formation of a
$Z$ boson condensate on the bubble wall, and that the competition
between macroscopic domains could enhance the difference
in free energy between the different sign condensates.

The most  clearly viable mechanisms elaborated to date work with
extensions of the standard model, the simplest such extensions being
those where there are extra Higgs fields. From a particle physics standpoint
this requires little justification and readily provides the
extra source of CP violation which seems necessary. Following
suggestions of Shaposhnikov
 \ref\shap{M. E. Shaposhnikov, JETP Lett {\bf 44}, 465 (1986); Nucl. Phys.
{\bf B287}, 757 (1987); Nucl. Phys. {\bf B299}, 797 (1988).}, and
McLerran \ref\lar{L. McLerran, Phys. Rev. Lett. {\bf 62}, 1075 (1989).},
Turok and Zadrozny
\ref\NTJTZa{N. Turok and J. Zadrozny, Phys.
 Rev. Lett. {\bf 65}, 2331 (1990).}
showed how this could work in multi-Higgs extensions of the standard model.
In these theories there is a CP odd Higgs field phase which changes
in a definite manner in bubble walls as the Higgs fields `roll'
from the unbroken to the broken symmetry vacuum. This phase
couples to the anomaly through a triangle diagram producing a biasing
of the anomalous processes
\ref\NTJTZb{N. Turok and J. Zadrozny,
Nuc. Phys. {\bf B 358}, 471 (1991).},
\ref\MSTV{L. McLerran, M. Shaposhnikov, N. Turok and M. Voloshin,
 Phys. Lett. { \bf  256B}, 451 (1991).}. As the bubble walls propagate
through the universe they leave behind a trail of baryons.

Cohen, Kaplan and Nelson (CKN) proposed two potentially more
efficient mechanisms in the same theories, `spontaneous'
baryogenesis
\ref\CKNsb{ A. Cohen, D. Kaplan and
A. Nelson, Phys. Lett. {\bf B263}, 86 (1991).}, and the `charge transport'
mechanism \ref\CKNct{A. Cohen, D. Kaplan and
   A. Nelson,  Nuc. Phys. {\bf B349}, 727 (1991).}.
The `spontaneous' mechanism
was an application of an earlier idea employing GUT baryon number
violation \ref\CKNsbold{ A. Cohen and  D. Kaplan, Phys. Lett. {\bf B199},
251 (1987);
 Nucl. Phys. {\bf B308}, 913 (1988).}, the charge transport mechanism
developed  an idea involving a lepton number violating phase
transition arranged to occur at the electroweak scale
 \ref\CKNctold{ A. Cohen, D. Kaplan and
   A. Nelson, Phys. Lett. {\bf B245}, 561 (1990).}.
Dine et. al.
\ref\DineHSS{M. Dine, P. Huet, R. Singleton and L. Susskind, Phys. Lett.
{\bf B257}, 351 (1991).} also gave a discussion of  `spontaneous'
baryogenesis employing electroweak baryon number violation in
general terms involving theories with a higher energy scale.
An interesting extension of these ideas is the work
of Comelli et. al. who showed that finite temperature
effects can cause the spontaneous violation of CP in the minimal
supersymmetric standard model. Bubbles of both CP types
nucleate, but any small explicit CP violation is greatly amplified
because it comes into the exponent of the bubble nucleation rate
\ref\cpr{ D. Comelli, M. Pietroni and A. Riotto, Nucl. Phys.
{\bf B412} (1994) 441.}.

CKN originally argued that `spontaneous' baryogenesis
would work well in multi-Higgs theories in the
`adiabatic' limit of slow and thick walls. A  phase redefinition of the
quark fields was used to identify a term in the Lagrangian
which acts like a  potential for fermionic hypercharge
on the bubble wall where the relative phase of the Higgs
fields is changing. CKN calculated the
local thermal equilibrium established
in this background, with certain `fast' perturbative processes
allowed to equilibrate,  and showed that it resulted in
a non-zero baryon number violation rate.
In  \ref\JPTi{M. Joyce, T. Prokopec and
N. Turok,
Phys. Lett. {\bf  B 339}(1994) 312 , hep-ph/9401351;
M. Joyce, in {\it Electroweak Physics and the
Early Universe}, eds. F. Freire and J. Romao, publ. Plenum, proceedings
of Sintra conference, March 1994, hep-ph/9406356.}
however
we argued that inappropriate constraints were imposed
on global conserved charges in this calculation and that the
 thermal equilibrium which is in fact approached  has
zero baryon violation. A correct calculation of this effect
involves finding the departure from this equilibrium
induced by the motion of the wall, determining
how the motion of the wall competes with transport
processes which tend to restore it to equilibrium.
In the limit of
perfect particle transport we showed that
the system approaches the thermal equilibrium
with zero baryon production.
Dine and Thomas
\ref\dt{M. Dine and S. Thomas,  Phys. Lett. {\bf B 328}, 73 (1994),
hep-ph/9401265.}
have also questioned whether it is correct to treat the system
as described by a fermionic hypercharge potential, pointing
out that this manifestly does not have the correct behaviour as
the Higgs vev  vanishes.

The second mechanism suggested by CKN, the
`charge transport' mechanism, was argued to apply  in the case
of thin bubble walls. The reflection of fermions off the
bubble walls in a CP violating way leads to the injection into
the unbroken phase of particle asymmetries which then
bias the baryon number violating processes in this region.
In \CKNctold\ this mechanism was applied to a heavy (4th generation)
majoron neutrino with  explicit $CP$ violation in Yukawa couplings.
Subsequently \CKNctold\ CKN showed how in a similar way the effect of $CP$
violation in the Higgs sector in a multi-Higgs extension of the
standard model could be mediated into the
unbroken phase by reflected top quarks, and
sizeable baryon asymmetries produced.
These calculations
were supposed to apply only to  thin walls because the
`quantum mechanical' reflection effect was taken to be
suppressed when the wall is thicker (i) than the quark
mean free path, because of scattering {\it or} (ii) than the
Compton wavelength, because of the validity in this
limit of the WKB approximation.

As pointed out by CKN this second mechanism has a great advantage over
previously suggested ones. It is `non-local' in the sense that
the CP violation and baryon number violation are separated
from one another. The effect of the CP violation on the wall
can be propagated away from the wall into the unbroken phase,
where the baryon number violating  processes are unsupressed.

In this and an accompanying paper
\ref\JPTlongb{M. Joyce, T. Prokopec and N. Turok,
Princeton preprint PUPT-1496,  hep-ph/9410282, Phys. Rev. {\bf D}, to appear.}
we will consider these ideas in a framework
which makes clear their relation to one another. Essentially
there is just one physical mechanism: on the
bubble walls a background
field is turned on which affects both
the dynamics and the interaction rates of the fermions.
Our calculations will apply both to the case
of a two doublet extension of the standard model and the case of
a $Z$ boson condensate in the standard model. We will
find that the property of non-locality can in fact be generalized
to the case of thick walls and used to greatly enhance the baryon
asymmetry in that limit. Corresponding to the reflection
in the thin wall case there is a classical force on
the wall which perturbs the particle densities on and in front of
the wall, as well as a `spontaneous' baryogenesis effect,
albeit in a guise consistent with the criticisms
of this mechanism in its original form. In the thin wall case we carry out
new analytic calculations which take into account a problem
in the original calculation pointed out in \JPTi.
 We develop a new analytic
diffusion formalism in which all
the distributions are determined dynamically. We concentrate
on the case of leptons emphasizing that they are very efficiently
transported into the unbroken phase and are likely to be best described by
this calculation for typical wall thicknesses.

The structure of the paper is as follows. In the next section
we discuss $B$ violation in the local thermal equilibrium
situation which is taken to pertain around and on the wall during
the phase transition and in particular we emphasize that
total left-handed fermion number should be
thought of as the driving force for $B$ violation.
In section 3 we discuss the general problem of the dynamics of
particles in the background of a CP violating bubble wall.
We discuss how the previously explored quantum mechanical
reflection has in fact a very non-trivial classical
limit. This  suggests a quite different approach to the
problem of modelling the response of the plasma to
the CP violating background when scattering on the wall
must be taken into account.
The rest of the paper is then concerned with
the thin wall regime and this other `classical'
calculation is deferred to the accompanying paper \JPTlongb.
In section 4 we calculate the  flux injected by reflection
off the wall. In Section 5 we describe our
calculations of the propagation of the injected flux into the
unbroken phase, and  give the derivation of a set of coupled equations for
the particle species which describes the diffusion and decay of
the injected asymmetries. In section 6 we solve these equations and
calculate the baryon asymmetries produced in various regimes.
In section 7 we discuss the issue of screening. While the diffusion  formalism
allows for a complete treatment of
screening, this addition
significantly complicates the analysis, and so for the most part we
ignore it. We attempt to justify this in
Section 7, where we give a qualitative argument that
including screening effects would alter our results
at most by a factor  of order unity.
In section 8 we calculate the
baryon to entropy ratio and discuss our results,
comparing the cases of injected quark and lepton
fluxes.
In the concluding section we summarize very briefly and
point out directions for further work.

\medskip

\centerline{\bf 2. Baryon Number Violation}

The departure from thermal equilibrium which is required for dynamical
baryogenesis occurs as the true vacuum bubbles sweep through the
false vacuum. We shall in this and the accompanying paper describe
this departure from equilibrium with space-time dependent
chemical potentials for each particle species. Whilst the perturbations
are not in general of this simple form, in many circumstances
as we shall argue in detail below, thermalisation towards
such a constrained local thermal equilibrium are rather efficient,
and this allows for a greatly simplified description of the baryogenesis
process.

The chemical potentials for the different particle species are fixed  by
the local values of the charges conserved by
`fast' processes which have time to equilibrate.
Other `slow' processes (such as baryon number violation)
remain out of equilibrium over the relevant timescales.
In these circumstances
the rate at which a charge $X$ conserved by the `fast' processes moves
towards its equilibrium value is given by
\eqn\xdot{\eqalign{\dot{X}=-\bar\Gamma {\Delta F \over  T} \Delta X }}
where $\bar\Gamma$ is the equilibrium rate for the process which violates
$X$, $\Delta F$ is the difference in free energy between the two states
(in which the process is in and out of equilibrium respectively) and
$\Delta X$ is the amount by which $X$ changes in the process. This equation
can be derived simply from detailed balance considerations.
Applying this to the case of $B$ violation one has simply
\eqn\bdoti{\eqalign{\dot{B}=-\bar\Gamma_s {\mu_B \over T} }}
where $\bar\Gamma_s$
is the sphaleron rate (per unit volume
per unit time)\footnote{$^\dagger$}{We will use
a bar over rates to denote rates per unit volume.}
and $\mu_B$ is the
chemical potential for $B$. CKN have phrased the problem in
these terms and set out in each of their scenarios to show
that $\mu_B$ is non-zero on or in front of the wall.

It is useful to
emphasize another form of this equation.
Consider a process $\nu_i n_i \leftrightarrow 0$
(in notation where, for example, $2n_1+n_2\leftrightarrow 3n_3$
has $\nu_1=2$, $\nu_2=1$ and $\nu_3=-3$) for which the rate
is $\Gamma$. If the particles are
in local thermal equilibrium with chemical potential $\mu_i$
then from \xdot\
\eqn\decayi{\eqalign{\dot{n_i}=
-{\bar\Gamma \over T}(\nu_i \Sigma_j \nu_j\mu_j).}}
Using the fact that
the sphaleron processes  are
$t_L t_L b_L \tau_L \leftrightarrow 0 $ and
$t_L b_L b_L \nu_L \leftrightarrow 0 $ (for a single fermion family
with obvious notations) we can write find
\eqn\bdotii{\eqalign{\dot{B}=-{\bar\Gamma_s \over 2T}
                  (3\mu_{t_L }+3\mu_{b_L }+\mu_{\tau_L}+\mu_{\nu_L }) }}
or, if there are $N_F$ families,
\eqn\bdotii{\eqalign{\dot{B}=-N_F{\bar \Gamma_s \over 2T}
    \Sigma_i (3\mu^i_{t_L }+3\mu^i_{b_L }+\mu^i_{\tau_L}+\mu^i_{\nu_L }) }}
where the sum is over the families. Here the
chemical potentials stand for the {\it difference} between particle
and anti-particle chemical potentials.
Equation \bdotii\ tells us that what we
must do in order to calculate $B$ violation is follow how the densities of
{\it left-handed fermions} and their anti-particles are perturbed.
In their work on charge transport baryogenesis CKN
have presented
hypercharge or a charge $X$ `orthogonal' to it as the
quantity driving baryogenesis. As discussed in \JPTi\
this emphasis is misplaced -
it is clear from \bdotii\ that one can have non-zero hypercharge
without any $B$ violation, and $B$ violation with
zero hypercharge. In the approach we describe in
these papers all charges will be determined dynamically
and we will always use the formula \bdotii\ directly.

In the case that the fermions are massless there is a simple relation
between the chemical potential and the perturbation to the number
density: for particles $\delta n_i = \mu_i T^2/12$, and the opposite
for antiparticles. In this case equation \bdotii\ becomes
\eqn\bdotiii{\eqalign{\dot{B}=-6N_F{\bar\Gamma_s \over T^3} (3B_L+L_L) }}
where $B_L$ and $L_L$ are the total left-handed baryon  and
lepton number densities respectively. The factor of 3 is just a
result of the definition of baryon number (i.e. ${1\over 3}$ per
quark) - the quantity in
brackets is just the total left-handed fermion number.

 We can use \bdotiii\  to comment on the role of
strong sphaleron processes. Giudice and
Shaposhnikov have pointed out
\ref\giud{G. Giudice and M. Shaposhnikov, Phys. Lett. {\bf B326}, 118 (1994).
}
 that if one treats these as
`fast' in
the constraint calculations carried out by CKN
(as they argue one should) one finds that the result is that
there is no $B$ violation. One must then go to the mass
induced corrections to \bdotiii\ to get a non-zero result. This
result has a simple explanation. In all these calculations $B$ and
$B-L_L$ are constrained to be zero. Also,  since strong sphalerons
couple right and
left-handed baryons as $t_L \overline{t}_R b_L \overline{b}_R....
\leftrightarrow 0$ it follows from \decayi\ that
the sum of the chemical potentials of left handed
baryons is equal to that of right-handed baryons if this
process is in thermal equilibrium. In the massless limit
this leads  to the relation $B_L=B_R$ which, together with the
constraints $B=0=L_L$ , immediately
gives $\dot{B}\propto 3 B_L +L_L =0$ from \bdotiii.
It is clear however that
if one injects net left-handed lepton number into the system that this
argument does not apply. We defer a full discussion of this
issue to the appropriate point below.

Now we turn to the analysis of how
CP violation on the wall perturbs
the left-handed fermion number on and in the
vicinity of the wall. This then
drives $B$ violation according to equation \bdotii.

\medskip

\centerline {\bf 3. Particle Dynamics on a CP-violating Wall}

We are interested in the effects of $CP$ violating condensates on
the bubble wall.
To date, the two simplest suggestions for this are
 the relative phase of two Higgs fields in a two-Higgs
theory  \NTJTZa, \NTJTZb,
or the longitudinal Z boson \snnt. Our analysis will include
both of these possibilities. The former case is conceptually simpler -
if there is explicit $CP$ violation in the Higgs potential,
the relative phase of the two Higgs fields is a CP odd field,
which changes in a definite manner as the Higgs fields change from
zero in the unbroken symmetry phase to their final nonzero values in
the broken symmetry phase \NTJTZa, \NTJTZb. The latter case relies
instead on an instability which may occur if the top quark mass is
sufficiently large \snnt.

Let us try first to define the relevant physical degrees of freedom.
We can use the $SU(2)$ symmetry to
pick a gauge in which the background classical solution provided
by the wall is
$\varphi_1={1\over \sqrt{2} }(0,v_1 e^{i\theta_1}))$,
$\varphi_2={1 \over \sqrt{2} } (0, v_2 e^{i\theta_2})$.
We assume
that at all stages during the phase transition electromagnetic
$U(1)$ symmetry is unbroken.
We could with a gauge choice set $\theta_1$ or $\theta_2$ to zero,
but it is instructive not to do so,
in order to clearly see how the (gauge invariant) $CP$-odd relative phase
$\theta =\theta_1-\theta_2$ emerges.
 It is clear  that since the relative
phase loses its meaning when the vevs vanish,
{\it all physical effects which depend on the relative phase
of the two Higgs fields must
vanish as either Higgs vev does}. We shall see that this condition
is met in a somewhat nontrivial manner.

The physical degrees of freedom
in the broken phase are fluctuations in the two vevs,
$\delta v_1$ and $\delta v_2$, in the
relative phase $\theta$,
and the charged Higgs fluctuations $\varphi^+$.
Diagonalising  the Higgs kinetic terms we find
\eqn\higgs{\eqalign{ |{\cal D}\varphi_1|^2
+|{\cal D} \varphi_2|^2 &=
 {1\over 2} {v_1^2 v_2^2 \over v^2}
\bigl (\partial_\mu(\theta_1 -\theta_2)\bigr )^2 + {1\over 2} v^2
\bigl[ {g\over 2}
 Z_\mu -\bigl ( {v_1^2 \partial_\mu \theta_1 + v_2^2 \partial_\mu
\theta_2 \over v^2}\bigr) \bigr]^2 \cr
&\equiv {1\over 2} {v_1^2 v_2^2 \over v^2}(\partial \theta)^2 +
{1\over 2} v^2 ({g\over 2}  Z_{GI})^2 \cr
}}
where $v^2=v_1^2+v_2^2$ and $g^2=g_1^2+g_2^2$.  We ignore
for now the kinetic terms for $v_1$ and $v_2$, the charged $W$ bosons and
the charged Higgs fluctuations $\varphi^+$ (we shall return to the latter
 below).
$Z_\mu$ is the usual (gauge variant)
expression in terms of $W_\mu^3$ and $B_\mu$.
Both $\theta$ and $Z_\mu^{GI}$ are gauge invariant, and we take them
as our definition of the $CP$ violating condensate field and
the $Z$ field respectively.

Assume for simplicity that only $\varphi_1$
couples to the fermions  via  Yukawa terms.
The phase of the Higgs field
$\theta_1$ can be removed
by performing
a $T^3$ (or hypercharge $Y$ - the two are equivalent since
$Q=T^3+Y$ is an unbroken exact symmetry) rotation on
the fermions,  at the cost of introducing
a coupling $\partial_\mu \theta_1 T^3$, i.e. a pure gauge
potential for  $T^3$  into the
fermion kinetic terms.  Let us see what effect this has.
We may combine the $\partial_\mu \theta_1 T^3$ potential with
the coupling to the $Z_\mu$ field to find
\eqn\redefi{\bar\psi i \gamma^\mu (\partial_\mu - ig_A \tilde{Z}_\mu
 \gamma^5)\psi - m\bar\psi\psi }
where $g_A= +g/4$ for up-type quarks
 and (left-handed) neutrino, and $g_A = -g/4$
for down-type quarks and leptons, and
\eqn\defnii{g_A \tilde{Z}_\mu= g_A Z_\mu^{GI} - {1\over 2} {v_2^2 \over v_1^2 +
v_2^2}
\partial^\mu \theta }
where  $g_A={1 \over 4} g$. The vector contribution
from the $T^3$ potential does not appear in  \redefi\
for the following reason.
We treat the wall as planar, and assume it has reached a static
configuration in the wall rest frame,
with the background scalar fields being functions only of $z$,
and  the field $\tilde{Z}_\mu = (0,0,0,Z(z))$ being  pure gauge.
We can then remove the vector term using the remaining
unbroken anomaly-free vector
symmetries: A hypercharge
rotation leaves a vector piece which couples
each fermion in proportion to $B-L$, which
can itself be removed by the appropriate
$B-L$ rotation. Precisely the charge $g_A$
as given in \redefi\ is the linear combination
 $[{1 \over 2}(T_3 -Y) +{1 \over 4}(B - L)]g$.
The remaining axial term  {\it cannot} be gauged away, and (as we shall
see) has a real physical effect.

Let us briefly mention how the
calculation is done in unitary gauge. In this gauge, the massive gauge
bosons are just
the original gauge bosons i.e. the Goldstone modes that would have
been `eaten' by the gauge bosons are set zero. This condition
is just that the gauge current corresponding to
each broken symmetry (with generator $T^a$ and corresponding gauge field
$W_\mu^a$), give by
 $\partial {\cal L} / \partial W^a_\mu$
be set zero at zero $W^a_\mu$. Then the Lagrangian ${\cal L}$ is quadratic in
$W^a_\mu$, and there is no need for any `eating' to occur.
In our case, this condition is just $i\varphi^* ( 1- \sigma_3) \partial_\mu
\varphi + {\rm h.c.} =0$,  or $v_1^2 \partial \theta_1=
- v_2^2 \partial \theta_2$.Then  $Z_\mu^{GI}=Z_\mu$,
and it is clear that the fermion rotation by $\theta_1 T^3$
induces a potential of exactly $\partial_\mu \theta_1 T^3$ into the
fermion kinetic terms. After dropping the vector coupling,
this agrees with \defnii.

In describing the rotations required to bring the Lagrangian
to the form of \redefi\ we did not consider the Higgs fluctuations.
As remarked by Dine and Thomas \dt, a $T^3$ rotation such as
the one we performed to remove the phases from the fermion mass terms
introduces some $\theta_1$ dependence into
the Higgs-fermion interaction terms.
To remove this one has to perform a  rotation on the
Higgs fields, which introduces some dependence
on $\partial_\mu \theta$ in the Higgs kinetic terms.
In the limit that the vevs vanish
this rotation is just a $T^3$ rotation of the Higgs
fields (or   $T^3 -Y$ when we perform the subsequent
hypercharge rotation to remove the vector term
in the fermionic Lagrangian),
and the induced term simply a coupling to
the field $\tilde Z_\mu$ which makes up
the remaining piece of the pure gauge potential.
The $g_A$ charges corresponding to the form
in \redefi\ are $0$ and $-g/2$ for the
charged and  neutral Higgs components respectively.
(so that $g_A$ charges which are just proportional
to a linear combination of $T^3 -Y$ and $B-L$
charges are conserved in any interaction).

Let us consider a little further the Higgs sector
when the vevs are not zero. The  $T^3$ rotation
on the fermions then induces a  $\theta_1$ dependence
in, for example, the charged Higgs-fermion interaction term
$\varphi^+ \overline{t_L} b_R$.
We now define the charged Higgs fluctuations
in unitary gauge.
With the notation $\varphi_1= (\varphi_1^+,
v_1/\sqrt{2})$ (i.e. absorbing the phase into $v_1$),
 and similarly for $\varphi_2$, one finds  the unitary gauge condition is
\eqn\unitary{
\varphi_1^{+*} \partial_\mu v_1+ \varphi_2^{+*} \partial_\mu v_2
= \partial_\mu \varphi_1^{+*} v_1 +\partial_\mu \varphi_2^{+*} v_2
}
Now  we write the kinetic terms for the
Higgs fluctuations:
\eqn\unitaryhiggs{\eqalign{
&K = |\partial_\mu \varphi_1^+|^2 +
 |\partial_\mu \varphi_2^+|^2 \propto   \cr
& |\partial_\mu \varphi_1^{+*} v_1
+\partial_\mu \varphi_2^{+*} v_2|^2
+|\partial_\mu \varphi_1^{+*} v_2
-\partial_\mu \varphi_2^{+*} v_1|^2 \cr
}}
We shall see how a potential for $T^3$ emerges in the Higgs sector
in a certain approximation, more restrictive than in the case of
the fermions.  If we drop derivatives
of $|v_1|$ and $|v_2|$, then we find upon substituting  \unitary\ for
the first term in \unitaryhiggs\ that it is quadratic in
$\partial_\mu \theta_1$ and $\partial_\mu \theta_2$.
Let us make the approximation of ignoring such terms.
If we now rotate the fermions by
$e^{i\theta_1 T^3}$ to remove the
phase from the fermion mass term, we must
rotate $\varphi_1^+$ by the same rotation to remove the phase
from the Higgs-fermion interactions, and $\varphi_2^+$ by the
same rotation to remove the phase from the Higgs-Higgs couplings.
Setting $\varphi_1^+ = f_1 e^{-i\theta_1}$
$\varphi_2^+ = f_2 e^{-i\theta_1}$, we find the second term in
\unitaryhiggs\ becomes
\eqn\higgstwo{\eqalign{
&K \propto
 |\partial_\mu f_1^{*} v_2
-\partial_\mu f_2^{*} v_1+i\partial_\mu \theta_1
(f_1^* v_2 - f_2^* v_1)|^2\simeq
| \partial_\mu \chi+i\chi\partial_\mu \theta_1 |^2
}}
where $\chi = f_1^* v_2 -f_2^* v_1$ is the physical
charged Higgs excitation. Thus, in this approximation,
one really can describe the effects of $\partial_\mu \theta_1$
as a $T^3$ potential for charged Higgs excitations (this was
originally argued to be true by Dine and Thomas \dt).

One other assumption we made in deriving \redefi\
was that only a single Higgs field coupled to the
fermions. If both Higgs fields couple to the
fermions we can carry out the same  procedure
to remove all the phases from the Yukawa terms
involving one Higgs field,
but are left then with the phase $\theta$ in
the Yukawa terms involving the second Higgs field.
There are alternative ways of writing this Lagrangian
- we can choose where to put the relative phase
but cannot get rid of it and make all the mass
terms real. In what follows we will always consider
cases where at most a single Higgs field couples
to any given fermion, so that in studying  the
dynamics of a given fermion in the CP violating
background we can always take its Lagrangian to
be of the form \redefi\ with a real mass term.

So what is the effect on the dynamics
of fermions of turning on an axially coupled pure gauge field in the
presence of a mass? And how are the distributions of the
particles in the plasma affected by this background?
We first consider the dynamics of a
free fermion described by the Lagrangian \redefi.
Because of translation invariance in the direction perpendicular
to $z$, the momentum $p_\perp$ is a constant of the motion.
With an appropriate Lorentz transformation, this may be set
equal to zero and the problem
becomes one dimensional.
With the ansatz $\psi \sim \exp -i\hat{E}t$, the Dirac equation can be broken
into
two equations, one coupling the first and third components
of $\psi$, and a second coupling the other two:
\eqn\diraciii{i\partial_z \xi_{\pm} =
 \pmatrix{ \hat{E} \pm g_A Z   & m \cr
           -m & -(\hat{E} \pm g_A Z ) \cr}
    \xi_{\pm}   \qquad \qquad  S^z = \pm {1\over 2} }
We are working here in the chiral representation to
follow the convention of CKN
in \CKNctold.
The energy $\hat{E}$ here equals $\sqrt{E^2-p_\perp^2}$
in the original Lorentz frame.
What \diraciii\ describes is the coupling of ingoing left-handed fermions
to outgoing right-handed ones and vice versa. For anti-particles the
same applies, except that the signs are switched i.e. left-handed particles
and their anti-particles (which are right-handed) see opposite signs for
the axial field. Since left-handed fermions and their anti-particles are
affected oppositely we have in principle a way of generating a disturbance
in left handed fermion number,
which should source baryon number as discussed in section 2.

This is precisely the effect identified by CKN in the `charge
transport' mechanism. Once one notes that the reflection from
the wall is CP violating one must simply calculate the injected fluxes
and determine how they bias baryon number violation.
Emphasis has
thus been placed on the calculation of quantum mechanical
reflection and transmission asymmetries.
We will discuss in detail in section 8 the criterion
for the validity of these calculations in which the
fermions are treated as free particles in their
interaction with the wall. Roughly the criterion is
that the thickness of the wall  $L_w$ be  much less
than the mean free path $\lambda_{mfp}$ of the fermions.
As the wall becomes thicker $L_w \sim \lambda_{mfp}$
strong scattering effects must be taken into account.
Because CP violating reflection has been understood, for
reasons which will be explained below, as a quantum
mechanical effect it has been assumed that it is washed out
in the limit $L_w >> \lambda_{mfp}$. In the next section
we discuss these reflection calculations, pointing out
that there is a non-trivial WKB result which leads to quite a different
conclusion and  points the way to a new calculation of the
thick wall case in which scattering of the fermions as they
cross the wall is taken into account.

{\bf 3.1 Two Calculations of Reflection Coefficients }

As discussed in section 2 the quantity of interest in
to baryon production is the difference in particle
minus anti-particle distributions.
We thus focus on calculating the difference in the reflection
probabilities of particles of a given chirality
and their antiparticles, since this is the quantity
that will enter the calculation (in section 4) of the
currects injected into the unbroken phase.

{\bf (i) Case 1 : Quantum Mechanical }

In Appendix A we derive an expression for the difference
$\cal{R}$ in the reflection coefficients
({\it i.e.\/} probabilities)
of particles and anti-particles.
For left-handed particles $L$ the result is
\eqn\rcoeffsi{ {\cal R}(p_z)\equiv R_{L \rightarrow R}-R_{\bar L\rightarrow
\bar R}=
           -{4 t (1-t^2)\over |m_f|} \int_{-\infty}^{\infty}
{\cal I}m [m(z) m_f^*] \cos (2 p_z z) \,dz}
where $m_f$ is $m(-\infty)$, the mass of the fermion in the broken
phase, and   $t={\rm tanh} \theta$ where ${\rm tanh } 2\theta
 = |m_f|/|p_z|$.
 Here $p_z$ denotes the value of the momentum at infinity
(in the unbroken phase). Henceforth we shall set
the phase of the fermion  mass  $m_f$
to be {\it zero\/} in the broken phase.
Equation \rcoeffsi\ is valid for any wall profile
and gives the leading term in an
expansion in  $L_w m_f $, where $L_w$ is the thickness of the wall.
Only when this expansion is valid
is there a range of incident
momenta which are both not totally reflected ($p_z > m_f$)
and also non-WKB ($p_z  <  L_w^{-1}$). It is because of this
that we
refer to it as the `quantum mechanical expansion'.

To see the effect of the cosine term in the integral
we evaluate
{\rcoeffsi\/} for an imaginary mass of a Gaussian
form : ${\cal I}m[m]=m_I={m_f\over \sqrt{\pi}}\exp -(m_H z)^2$.
The result for the reflection coefficient is then:
\eqn\reflection{{\cal R}=
       4t(1-t^2){m_f\over m_H} \Theta_{CP}
          e^{-\left({p_z\over m_H}\right)^2}\,,}
where $\Theta_{CP}$ is defined by
$\int {\cal I}m\,[m] dz={m_f \over m_H}\Theta_{CP}$.
\reflection\  illustrates
that the effect is quantum mechanical, as it is
exponentially suppressed in the WKB   limit, and
in the limit    $|p_z| L_w < 1$    clearly involves
a non-local coherence effect across the wall
in which the particle and anti-particle
pick up different phases as they propagate.

This is the effect identified by CKN when they
calculated the reflection coefficients numerically
for a definite wall profile, and also in
\ref\funakubo{K. Funakubo, A. Kakuto, S. Otsuki, K. Takenaga
and F. Toyoda, Phys. Rev. {\bf D50}, 1105 (1994);
preprints hep-ph/9405422, hep-ph/9407207, hep-ph/9503495}.
They found strong suppression of reflection for momenta larger
than the inverse wall thickness and concluded correctly
that they were seeing a  phase coherence effect
which would be strongly suppressed for all momenta
when the effects of scatterings on the bubble wall
are accounted for. Consideration
of the WKB  limit will now show that this analysis misses
an important point.

{\bf (ii) Case 2 : WKB Regime}

We now consider the limit in which the functions $m$ and $Z$
vary slowly in comparison to the Compton wavelength of the
fermion. The WKB solutions to \diraciii\
then describe excitations which
may be thought of as particles with definite momentum and position.
They are described by the dispersion relation
\eqn\wkbiv{E =
\left [{p_\perp^2+
(\sqrt{p_z^2+m^2} \mp g_A Z)^2}\right ]^{1/2} \qquad \qquad S^z=\pm {1\over 2}}
where $S^z$ refers to the spin of the particle in the
frame where $p_\perp$ vanishes. For particles incident from the unbroken
phase this is equivalent to chirality. This equation only
makes sense for $m> |g_A Z|$, and we assume this holds
everywhere. If
this condition is violated, `positive energy' solutions can
mix with `negative energy' ones and one expects particle creation to occur
(as in the Klein paradox). We shall not consider this possibility here.

A particle incident from the unbroken phase
with momentum $p_z(\infty) = p_\infty$
is reflected if
\eqn\wkbv{|p_\infty| \; < \; m(z)\mp g_A Z(z) }
holds for any $z$. From \wkbv\
we can read off the difference in the reflection coefficients:
\eqn\wkbref{  R_{L \rightarrow R}-R_{ \bar L \rightarrow \bar R }=
    \cases { 1 , & $|p_\infty| \in \bigl [ m_f, \max [m(z) + g_AZ(z)]\bigr ]$
 \cr
             0 & {\rm otherwise}\cr}    }
(where we have assumed $g_A Z(z)$ is positive),
ignoring barrier penetration effects.

What this analysis reveals is that
high $p_z$ (i.e. WKB) particles see the axial field on the wall
as producing an extra potential superimposed on the
real mass barrier.
This extra potential has an opposite sign for particles and anti-particles.
If the shape of the extra potential is monotonic like that
shown in Figure 1 this potential does
not produce a difference in reflection coefficients for the
`classical' particles and their anti-particles. The only
interesting reflection will come at low `quantum mechanical'
momenta for which the different phases picked up by
particles and antiparticles
moving in this potential  causes differential reflection.
However if the
potential looks more like that in Figure 2 (with a bump) there will
be a difference in reflection coefficients of particles and anti-particles
even in the WKB limit.
Thus we see that the exponential
suppression we observed in \reflection\
was a result of the particular
ansatz we took for the wall profile. Similar monotonic ansatzes have been
assumed by CKN and other authors  \funakubo.

Besides revealing
a sensitivity to the ansatz in the calculation of the reflection
coefficient, in particular that there can be non-trivial
reflection of WKB particles, this discussion is of relevance
to the thick wall case. The effect being treated
was expected to be suppressed when scattering on the wall becomes
important since  it relies on a phase coherence of the
particles across the wall. However the effect does not arise
from such a phase coherence in the case
of WKB momenta particles.
As we have argued the
picture appropriate to these momenta is of a (local)
classical potential
which is simply different for particles and anti-particles.
What do we expect such a potential to give rise to?
Consider the effect of turning on an electromagnetic
potential in a small region of a plasma. Given a sufficient amount of
time the potential will be screened - it will draw in charge - and
settle down to a new thermal equilibrium in which there is
a net overdensity of every particle in proportion to its electrical
charge times the local value of the potential.
If we consider the case in which the region moves slowly
with velocity $v$ the system will remain approximately in this thermal
equilibrium
but with deviations which vanish as
$v \rightarrow 0$.
We have just argued that the particles and
anti-particles see an opposite potential and we thus expect that
the deviations induced as the region - in this case the
bubble wall - moves will reflect this difference. This should be true
irrespective of whether the potential is monotonic or not. Just as
in the case of an electromagnetic potential where the process
of establishing the approximate equilibrium involves the electromagnetic
force pulling in particles which then scatter in the potential and
reach local equilibrium, there should be a force playing an analagous role.
This force will be a CP violating force which will source perturbations
in front of the wall much as the reflected flux does in the thin
wall case.

What is the force a particle feels moving on the wall? To answer
this we return to the dispersion relation \wkbiv. The velocity
of the WKB particle with energy $E$ is the group velocity
\eqn\group{v_z= {\partial E \over \partial p_z}={\sqrt{E^2-p_\perp^2}
\over E}{p_z \over \sqrt{E^2 - p_\perp^2}\pm g_A Z }\qquad S^z=\pm{1\over 2}}
where $p_z$ is specified by \wkbiv. A simple measure of the effect of
the background is the corresponding acceleration, which
may be calculated using $\dot{p}_z = -\partial_z E$:
\eqn\accel{ {dv_z \over dt}= - {1 \over 2 }{(m^2)' \over E^2}
\pm { (g_AZm^2)' \over E^2 \sqrt{E^2 - p_\perp^2}}+ o((g_A Z)^2)}
with $E$ and $p_\perp$ constants of the motion.
The $CP$ violating effect goes to zero as the mass squared does,
consistent with the fact that $Z$ may be gauged away in this limit.

In what limit is this force associated with the background likely
to be important? We turn to this question in the accompanying
paper \JPTlongb, where we develop a new formalism to describe
the effect. An accurate treatment of the `nonlocal'
classical effect
becomes possible for thick, slow walls, where a fluid approximation
may be used in order to take particle scattering on the wall
into account.
But in the remainder of this paper we turn
to the consideration of the opposite `thin wall' quantum mechanical  limit,
taking the fermions to propagate as free particles in
the background provided by the wall.

\medskip

\centerline{ \bf 4. Injected Fluxes in the Thin Wall Regime}

We first make a few comments about the physical parameters which determine the
applicability of the calculations we are undertaking.
In these papers we assume  that the phase transition is first
order and proceeds by bubble nucleation. This is supported by
detailed study of the perturbative effective potential at least
for rather light Higgs ($m_H<70\, GeV$) \ref\ArnoldFodor{
 P. Arnold and O. Espinosa,  Phys. Rev. {\bf D47}, 3546 (1993);
W. Buchm\"{u}ller Z. Fodor, and A. Hebecker, Phys. Lett. {\bf B331},
131 (1994);
W. Buchm\"{u}ller, Z. Fodor, T. Helbig and  D. Walliser,
Ann. Phys. {\bf 234}, 260 (1994), hep-ph/9303251.}  and,
more recently, by non-perturbative studies
\nonpert.
The parameters which are
crucial to an accurate determination of the baryon asymmetry
are the wall thickness $L_w$ and the
wall velocity $v_w$.
Estimates of the bubble wall thickness from
perturbative calculations in the standard model
\ref\wall{N. Turok, Phys. Rev. Lett. {\bf 68}, 1803 (1992);
M. Dine, R. Leigh, P. Huet, A. Linde and D. Linde,
Phys. Rev. {\bf D46}, 550 (1992);
B-H. Liu, L. McLerran and N. Turok, Phys. Rev. {\bf D46}, 2668 (1992);
P. Huet, K. Kajantie, R. G. Leigh, B.-H. Liu and
 L. McLerran,  Phys. Rev. {\bf D48}, 2477(1993).}
indicate  $L_w \approx  2m_H^{-1}\approx {10-40 \over T}$.
and $ v_w \sim 0.1 -1$.
A recent detailed study by one of us (T.P) and G. Moore
\ref\MooreP{T. Prokopec and G. Moore, Phys. Rev. Lett. {75},777(1995)
and  PUP-TH-1544, LANCS-TH/9517 (1995).}
in the minimal standard model indicates that for
Higgs masses in the range  $30-70$  GeV or so,  $v_w\sim 0.4$
and  $L_w \sim 25/T$.
If non-perturbative effects play a significant role
this result could be significantly altered; a
non-perturbative condensate makes the phase transition stronger
with more super-cooling and hence larger $v_w$. Likewise
in two-Higgs theories larger or smaller $v_w$'s may be possible
in different regions of parameter space.


The other parameters which are crucial to accurate calculations
are those characterizing the particles in the hot plasma.
One of our objectives in this work has been to carry
out detailed calculations of diffusion constants
and decay rates rather than give rough estimates of these
parameters. This is particularly important because the emphasis
of our work is on transport and the transport properties of
quarks and leptons are quite different.
An estimate of the length scale over which we can treat the fermion
as free is specified by the fermion damping rate $\gamma$
in a hot plasma \ref\braaten{E. Braaten and R. Pisarski,
Phys. Rev. {\bf D46}, 1829 (1992);
Phys. Rev. {\bf D47}, 5589 (1992);
Phys. Rev. {\bf D42}, 2156 (1990).}, {\ghop\/},{\hs\/}.
For the slow quarks and leptons ($p\leq gT$) these are
\eqn\damp{ \gamma_q  \approx 2\alpha_s T \approx  {T \over 3.5} \qquad
          \gamma_l  \approx \alpha_w T \approx  {T \over 30}  \qquad
\gamma_r  \approx {3 \over 2}
\alpha_w \tan^2 \theta_w T \approx  {T \over 70}   }
at $100$GeV, where $\theta_w$ is the Weinberg angle, and the subscripts
indicate quarks, left-handed leptons and right-handed leptons.
For hard thermal particles ($p\sim T$) even though the damping rates are
logarithmically enhanced: $\gamma\sim g^2\ln (1/g^2)$, the numerical value does
not differ by much: $\gamma_q\sim 4\alpha_s T/3$,
$\gamma_{l}\sim \alpha_w T$.  Comparing these
with $L_w$ we see
that the uncertainty in $L_w$  translates to an uncertainty about
whether the fermions should be treated as free or interacting particles
on the wall.
It seems most likely with current understanding of the relevant
parameters that leptons
may be well described by a free particle treatment on the wall,
but for quarks an interacting fluid approximation is needed.
In this paper we shall consider the former case, in
the companion paper \JPTlongb\ the latter.

In the context of these remarks about the likely difference between
the quarks and leptons we note an interesting point related to the finite
temperature properties
of a two Higgs doublet model, which we believe has been overlooked
in the literature. The ratio of quark and lepton masses is not in
general determined by their zero temperature ratio.
For example the ratio of the {\it top\/} quark mass
to the $\tau$-lepton mass at finite temperature in the models with the
couplings described above is
\eqn\massratio{\eqalign{{m_t(T)\over m_\tau(T)}={y_t\over y_\tau}{v_1(T)\over
v_2(T)}=
               {y_t\over y_\tau}{v_1(0)\over v_2(0)}
               {v_1(T)/ v_1(0)\over v_2(T)/ v_2(0)}=
               {m_t(0)\over m_\tau(0)}{v_1(T)/ v_1(0)\over v_2(T)/ v_2(0)}\,.}
}
There is no reason why the ratio of the zero temperature masses should not
arise from a tuning of the {\it vev}s in the potential rather than from a
tuning of the Yukawa couplings. If there is such a tuning we do not
expect it to be respected by finite temperature corrections to the potential.
Rather we expect all the finite temperature {\it vev}s to be of order $T$.
Thus it is the ratio of the Yukawa couplings which is important and
these can be large or small, consistently with present experimental data.

We now turn to the calculation of the injected flux in the thin wall
regime which we will use in the next section to calculate the
departure from thermal equilibrium in the unbroken phase.
In order to  get an expression for the asymmetry  injected into
the unbroken phase, we need to include both reflection
from and tranmission through the barrier. One can use
CPT invariance to find the following relations between
the reflection and tranmission coefficients \CKNctold
\eqn\reftrans{R_{R\rightarrow L}=R_{\bar L \rightarrow \bar R}=
1-T_{L\rightarrow L}=1-T_{\bar R\rightarrow \bar R} }
where the transmission coefficients are for particles
incident from the broken phase.
Integrating the reflection and transmission
coefficients against the incident flux, and using
\reftrans\ to eliminate the transmission coefficients,
we arrive at an expression
for the flux of injected right handed particles
\eqn\fluxgen{J_i^R= \int_{p_z<0} {d^3 p \over (2 \pi)^3} { |p_z| \over E}
(f_{\leftarrow}^L-f_{\rightarrow}^R) {\cal R}(p_z) }
where $f_{\leftarrow}^L$  and  ($f_{\rightarrow}^R$) are the
free particle phase-space densities (in the wall frame)
of left-moving left-handed (L)
and right-moving right-handed (R) particles,
respectively.
When the wall is at rest the term $f_{\leftarrow}^L-f_{\rightarrow}^R$
cancels so that there is no net reflection.
Taking this term to linear order in $v_w$, and
using the coefficients in
\rcoeffsi, calculating again to leading order in  ${m_f \over m_H}$, and
also to leading order in $m_H \over T$, we find
\eqn\rfluxqmi{\eqalign{J_i^L  = -J_i^R
 = {{v_w\over {4\pi^2}}  m_f^2 m_H\Theta_{CP}
        \,. }}}
For the reflection coefficient we have taken
\eqn\reflapprox{{\cal{R}}(p_z)= \cases{ {2 m_f \over p_z }{m_f \over m_H}
\Theta_{CP} & $m_f < p_z < m_H$ \cr 0 & otherwise \cr } }
In imposing the upper cut-off we assume that there
is a sharp WKB suppression like that in \reflection.
The approximation $2t \approx tanh 2\theta = {m_f \over p_z}$
is good to leading order in ${m_f \over m_H}$
since the integrals are dominated by the larger momenta
because of phase space factors.

In the WKB case, using \wkbref,  we obtain
\eqn\wkbcurrent{J_i^L=-J_i^R={v_w\over 4\pi^2}m_f^2m_H \Theta_{CP}\eta}
where we took $g_A Z_{\rm max} =m_H\Theta_{CP}$.
The parameter $\eta$ parametrises our  ignorance
about the exact profile and penetrability
of the barrier. For a thick (impenetrable)
barrier with a prominent bump such that
${\rm max}[ m(z)+g_A Z]=m_f+g_A Z_{\rm max}$, $ \eta=1$. In general $\eta<1$.
All masses in these formulae are the relevant finite temperature masses.

What is the correct ansatz for the wall profile?
Although monotonic ansatzes of the type previously assumed
seem plausible they are no more than that. This point
is well illustrated by the calculations in
\ref\turoknasser{S. Nasser and  N. Turok, in preparation.}
in which it is shown that in the standard model the backreaction
of the CP violating reflection can cause an instability
to the formation of a $Z$ condensate. Without a detailed calculation of
these effects it remains unclear exactly what the wall profile is.

\medskip

\centerline{\bf 5. Propagation of Injected Particle Asymmetries}

We now consider the problem of how the flux calculated to be injected
in the first two cases above propagates in the unbroken phase and
drives $B$ violation. The average velocity $v_i$ relative to the wall of the
injected flux
can be written as
\eqn\vigen{ v_i={{\int_{p_z<0} {d^3 p \over (2 \pi)^3} { p_z^2 \over E^2}
(f_{\leftarrow}^L-f_{\rightarrow}^R) {\cal{R}}(p_z) \over
\int_{p_z<0} {d^3 p \over (2 \pi)^3} { |p_z| \over E}
(f_{\leftarrow}^L-f_{\rightarrow}^R) {\cal{R}}(p_z) } } }
Using \rcoeffsi\ and the same approximations
as for the calculation of the flux
(to leading order in $v_w$, $m_f /m_H$ and $m_H / T$)
we find
\eqn\vinj{ v_i ={1 \over 4ln2} {m_H \over T}.   }
For the WKB case, using \wkbref\ and \fluxgen, the result is
\eqn\vinjwkb{\eqalign{ v_i={\pi^2 \over 18 \zeta_3} {m_f \over T}     }}
where $\zeta_3\simeq 1.202$ is a Riemann zeta function.
The result in both cases is  roughly
the ratio of the momentum in the direction of
the wall of the typical reflected particle to
the average thermal momentum in that direction.
Note that $v_i$ is
independent of the wall velocity.
The particles in this flux propagate away from the wall until they
scatter. If the wall is not moving too fast they will then
thermalize
and diffuse until they are overtaken by the wall. Any given particle
can also decay into others through decay processes and these must
be taken into account.

For what range of wall velocity $v_w$
do particles have time to thermalize before they are recaptured by the wall?
Suppose that $\tau$ is the mean time for a particle's velocity
to be randomized.
We take $\tau$ to be both a thermalization time and
the step time in an isotropic random walk which the particle executes once it
thermalizes. Then we can estimate the mean distance a particle
moves away in the direction of the wall motion in time $t$ to be
$\sqrt { { t\tau \over 3} }$. Equating this to the distance $v_wt$ the
wall moves in the same time we see that the
the ratio of the `injection time' $\tau$
to the `diffusion time' $t$ is approximately $3v_w^2$.
We will restict ourselves to the case when this ratio is
small i.e.
\eqn\vsound{ v_w << v_s={1\over \sqrt {3}} }
where $v_s$ is the speed of sound
in the relativistic plasma. We can then
model the problem analytically with equations
describing the diffusion and decay of a particle density
perturbation sourced by an injected flux localized in a small
region around the wall.

{\bf 5.1 Calculation of Diffusion Constants}

We present our calculation of the relevant diffusion constants
in Appendix C. We generalize the standard
treatment of this problem given in
\ref\Lif{E. M. Lifshitz and
L. P. Pitaevskii, {\it Physical Kinetics\/}, Pergamon Press (1979).}.
We assume the local
distribution function $f(x,p)$ to be specified by a space
dependent chemical potential plus a perturbation which we
fix by balancing the contribution to the change in $f$ due
to the gradient in the chemical potential against the collision
term in Boltzmann's Equation. For the dominant gauge boson
exchange diagrams we then arrive at expressions for the diffusion
constant $D$ which defines the relation between the diffusion
current $\vec J_d$ and the gradient driving it:
\eqn\dflux{\eqalign{ {\vec{J}_d}= - D {\vec{\nabla}}n   }}
where n is the number density.
We will give our numerical results at the appropriate point below.

{\bf 5.2 Derivation of Propagation Equations}

We now use the continuity equation to derive the equations which
describe the propagation into the unbroken phase.
This gives
\eqn\cnty{\eqalign{ { dn_i \over dt}= { \partial n_i \over \partial t}
              + \vec{\nabla}\cdot\vec{J_i} = -\sum_A {\bar \Gamma_A\over T}
                   \nu_i(\sum_j \nu_j^A \mu_j).    }}
The decay term is derived using \decayi\ and the sum over $A$ is a sum
over all the decay channels of particle $i$.
Taking the leading (massless) term we can express this in terms of the
particle densities. In Appendix D
we outline the calculation of the specific decay rates which we
will need below.

Now making use of \dflux\ and \cnty\ and treating the wall as providing a flux
$ \vec{J}^{\, inj}_i $ so that
$\vec {J}_i= -D_i \vec{\nabla}n_i +\vec{J}_i^{\, inj}$,
we find
\eqn\cntyi{\eqalign{ D_i\partial_z^2 n_i - \partial_t n_i -\sum_A {\bar
\Gamma_A\over T}
           \nu_i(\sum_j \nu_j^A \mu_j)= \partial_z {J_z}^{\, inj}_i . }}
The wall can be treated in this way since it conserves particle number,
simply sending equal and opposite particle number in opposite directions.
It can thus be described as producing a `blip' in the flux, and taking it
to be constant in the wall frame, we model it as
\eqn\injflux{\eqalign{  J^{inj}_i(z) =  \xi_i  J^0_i  \delta(z-v_wt)  }}
where $J^0_i$ is the net reflected flux for each species
$i$, given for example for right handed particles
by equation \fluxgen. We define a parameter here called
 $\xi_i$ which defines
the persistence length of the current in the vicinity of the wall,
and we approximate the injected current with a delta function.
This is
reasonable if the injected current thermalizes in a time
$\tau_{th}$ short in comparison to the time a particle
spends diffusing before being recaptured by the wall
- which is just the criterion \vsound.

Our  ignorance about
how the injected flux thermalizes is parametrized by $\xi_i$.
The uncertainty in this parameter is
unfortunately intrinsic to the analytic approximation we are using
in taking the departure from thermal equilibrium to be
modelled by a chemical potential, a form
to which the injected flux does not in general conform.

Nevertheless an estimate of $\xi_i$ can be made as follows.
For a diffusing particle with diffusion constant $D$
the velocity randomization time $\tau$ is $\sim 6D$.
This can be obtained from the following simple consideration.
The average distance a particle with a velocity
randomization time $\tau$ diffuses
in a given direction in time $t$ is $\langle z^2\rangle =
{1 \over 3 } t \tau $.
On the other hand the solution to diffusion equation $\partial_t n= D
\nabla^2 n$  with a point like source:
$n(\vec{r}, t) \sim  {1 \over t^{3/2} } e^{-r^2 / 4Dt}$ specifies:
$\langle z^2\rangle=\langle r^2\rangle/3= 2Dt$. A simple comparison of the
two gives $\tau\sim 6D$.
Using this decay
time for a flux injected at velocity $v_i$, we estimate
$\xi \sim 6D_i v_i$.
$\xi$ can also be taken to include other
suppressions of the injected flux which may be relevant. We will
return to a discussion of this question in section 8.

Finally we look for solutions which are stationary in the rest frame of
the wall, i.e. of the form $n(z,t) \equiv n(z-v_wt)$. The
equations are then
\eqn\dde{\eqalign{ D_i n_i'' + v_w  n_i' -\sum_A \Gamma_A
           (\sum_j k_j^A n_j)= \xi_i J^0_i \delta'(z-v_wt) }}
where $k^A_i$ is the factor which results when we express the decay terms
in terms of the particle densities and $\Gamma_A={12 \over T^3}\bar \Gamma_A$.
The problem now becomes that of determining the baryon density behind
the wall (in the broken phase where we take the electroweak
sphaleron processes to be turned off)
when a flux of some species is injected in front of  the wall.

\medskip

\centerline{\bf 6. Solution of the Propagation Equations}

To understand the solution of these equations let us first
consider the case in which a chiral tau lepton flux is injected
by the wall. The only decay processes we will consider
are the Higgs mediated processes shown in Figure 3 and
the electroweak sphaleron processes. In the unbroken phase
there are no other processes which cause a chiral lepton
to decay; in the broken phase there are other vev-suppressed
processes but we will neglect these, taking them to be slow in
a sense which will become more precise through our analysis.
For the moment we will assume
that the electroweak sphalerons are too slow to significantly
alter the densities
in front of the wall and thus that they too can be neglected,
except for their role in making baryons.
The Higgs processes bring about a coupling between the  tau leptons
and several other species ( Higgs particles, weak gauge bosons, quarks).
For simplicity let
us assume that these induced densities in species other than tau
leptons can be neglected.
We will later relax these assumptions and treat the full
set of coupled equations.
We then have
\eqn\leptonsi{\eqalign{D_L L_L''+v_wL_L' -\Gamma_{LR}(aL_L-bL_R) &= \xi_L J_0
                         \delta' \cr
    D_R L_R''+v_wL_R' +\Gamma_{LR}(aL_L-bL_R) &= -\xi_R J_0
                         \delta' \cr     }}
where we keep $a={1\over2}$ and $b=1$ as variables for heuristic purposes.
Left handed particles come as isospin doublets, but only one particle
participates in each interaction. This explains $k_L\equiv a=1/2$.
The problem of solving \leptonsi\ reduces to
 the corresponding homogeneous equations
subject to the boundary conditions
\eqn\bcsi{\eqalign{ D_L L_L |{^+_-}&= \xi_L J_0
              \qquad D_L L_L' + v_w L_L |{^+_-}= 0 \cr
                     D_R L_R |{^+_-}&= -\xi_R J_0
              \qquad D_R L_R' + v_w L_R |{^+_-}= 0 \cr}}
which are derived by integrating up the equations through $z=0$,
imposing the conditions that the $L_L$ and $L_R$
are at most step-like discontinuous across the wall
and that they are zero at $+ \infty$
(in the unbroken phase).

Substituting the ansatz $e^{-\mu z}$ into \leptonsi\ we find that
the exponents are the roots of the equation
\eqn\quartic{\eqalign{D_LD_R\mu^4 - v_w(D_L + D_R)\mu^3
    +[v_w^2-\Gamma_{LR}(aD_R+bD_L )] \mu^2+v_w(a+b)\Gamma_{LR} \mu = 0    .}}
This gives a constant solution and a cubic which has two real positive roots
and one negative root. Requiring that the solutions be zero
at $+ \infty$
and finite at $- \infty$
we have eight constants to determine in the
solutions:
\eqn\solutioni{\eqalign{ L_L &=  L_1 e^{-\mu_1 z}+ L_2e^{-\mu_2 z} \qquad z>0
\cr
                             &=  L_3 e^{-\mu_3 z}+ L_4 \qquad z<0 \cr
                         L_R &=  R_1 e^{-\mu_1 z}+ R_2e^{-\mu_2 z} \qquad z>0
\cr
                             &=  R_3 e^{-\mu_3 z}+ R_4 \qquad z<0 \cr}}

where $\mu_3$ is negative.
The conditions \bcsi\ fix only four of these constants. The
additional  condition required is found by adding the equations and
integrating. This tells us that
\eqn\bcsii{\eqalign{ D_L L_L' + D_R L_R' +v_wL_L + v_w L_R     }}
is a constant of motion. By fixing it everywhere we can determine all
the constants.

This conserved quantity \bcsii\ is
simply the dynamical version of lepton number conservation which is built
into the equations. More generally there is exactly one such relation
for every quantity conserved by the decay processes we include in
\dde.
$B-L$ conservation,  for example,  will take the form
\eqn\bminusl{\eqalign{D_{q_L} B_L'+ D_{b_R} B_R' +D_{t_R} T_R'
-D_{l_L} L_L'-D_{l_R} L_R'+v_w(B-L) = 0}}
where $D_{q_L}$ is the diffusion constant for left-handed quarks etc.
If all the diffusion constants are equal and the injected $B-L$ is zero then
$B-L$ will be zero everywhere. However since in general the various species
diffuse differently this will result in non-zero $B-L$ locally, consistent with
the constraint of global $B-L$ conservation embodied in \bminusl.

If we can solve the equations \leptonsi\ to find $L_L$ we can then use
\bdotiii\
to find the induced baryon number. Taking the baryons to be thermalized they
appear as a flux $-v_w B$ in the wall frame and therefore
\eqn\bdotiv{\eqalign{\dot{B}= -v_w \partial_z B=-6N_F{\bar\Gamma_s \over T^3}
L_L
                      \equiv -{\Gamma}_s L_L. }}
This equation is simply the limit of the appropriate diffusion/decay
equation for baryon number which results when we use the fact that
the diffusion constant of the quarks is very much less than the
diffusion constants of the leptons, which source the equation.
Integrating up $L_L$ as in \solutioni\ we find the baryon number
on the wall to be
\eqn\bzeroi{\eqalign{B(0)= - {{ \Gamma}_s \over v_w}({L_1 \over \mu_1}+{L_2
\over \mu_2}) }}
Thus we need simply to determine $L_1$ and $L_2$ by solving the cubic
equation for the roots and the boundary problem.

We can solve \quartic\ approximately in two different limits:

{\bf Case 1.} $v_w^2 >> \Gamma_{LR}(aD_R+bD_L) $.

In this limit the roots are
\eqn\rootsi{\eqalign{\mu_3=-(a+b){\Gamma_{LR} \over v_w} \qquad
                     \mu_1={ v_w \over D_L}+ {a \Gamma_{LR}\over v_w}\qquad
                     \mu_2={ v_w \over D_R}+ {b\Gamma_{LR}\over v_w} }}
and, using the boundary conditions \bcsi\ and \bcsii, we find
\eqn\solni{\eqalign{L_1&={\xi_L \over D_L}J_0 \qquad \qquad R_2=-{\xi_R \over
D_R}J_0 \cr
                    L_2&={b\Gamma_{LR} D_R \over v_w^2}R_2 \qquad
                     R_1={a\Gamma_{LR} D_L \over v_w^2}L_1 .   \cr  }}
 From \bzeroi\ we then have
\eqn\bzeroii{\eqalign{B(0)= - ({\xi_L \over D_L} J_0)
 [{{\Gamma}_s D_L \over v_w^2} -{ {\Gamma}_s (bD_R+aD_L)\over v_w^2} {
\Gamma_{LR} D_R\over v_w^2}]. }}
We have taken ${\xi_L / D_L}={\xi_R / D_R}$, assuming
$\xi_i \sim D_i$ from our estimate above.
The first factor in \bzeroii\
is the amplitude of the diffusing flux, down by
${\xi_L / D_L}$ relative to the injected flux amplitude due to
back diffusion across the wall. The expression inside the brackets is
a conversion factor of this injected flux to baryons.

{ \bf Case 2.} $v_w^2 << \Gamma_{LR}(aD_R+bD_L) $

Now the roots are
\eqn\rootsii{\eqalign{\mu_1= { a+b \over  aD_R + b D_L}v_w \qquad
                     \mu_2=-\mu_3= \sqrt{\Gamma_{LR}(aD_L^{-1} + b D_R^{-1})}
}}
and the solutions
\eqn\solnii{\eqalign{L_1&={b \over a}R_1 \qquad R_1=
                  -{aD_R \over aD_R+bD_L}({\xi_R \over D_R}J_0)
                  +{aD_L \over aD_R+bD_L}({\xi_L \over D_L}J_0) \cr
                    L_2&=
                  {bD_R \over aD_R+bD_L}({\xi_R \over D_R}J_0)
                +{aD_R \over aD_R+bD_L}({\xi_L \over D_L}J_0)
                    \qquad R_2=-{D_L \over D_R}L_2    \cr }}
to leading order in both ${v_w^2 / \Gamma_{LR}D_R}$ and ${v_w^2 /
\Gamma_{LR}D_L}$.
\rootsii\ and \solnii\ give
\eqn\bzeroiii{\eqalign{B(0)=  ({\xi_L \over D_L} J_0)
                { {\Gamma}_s ( D_R -  D_L) \over v_w^2}{b \over a+b} .}}

These two limits of these equations have a simple explanation.
The mean time a particle with diffusion constant $D$ spends in
the unbroken phase before it is captured by the wall is
$\sim {D / v_w^2}$. If, as in case 1, this is long in
comparison with the decay time
$\Gamma^{-1}$ then the solution describes the two injected fluxes
diffusing with a perturbation to each of order
${\Gamma_{LR} D/ v_w^2}$ which tells us
how the two species are sourced by one another through the decay process.
Likewise since $B$ violation has been assumed to be a
slow process and there is no injected $B$ the conversion factor is
for the same reason ${{\Gamma}_{s} D/ v_w^2}$.
\bzeroii\ therefore simply
shows how $B$ is sourced directly by the injected $L_L$ and
indirectly by the decay of the injected $L_R$ into $L_L$.
The precise coefficients can be read off from the decay term.
We note that the signs of the contributions are opposite and
therefore that there can be (for certain decay and diffusion constants)
a critical wall velocity $v_w$ for which they will precisely cancel.

The second case above is that in which the particles have time to
decay before they are caught by the wall. In the $\mu_1$ solution
which extends furthest into the unbroken phase the
left and right-handed leptons are in abundances such that
the decay process is approximately (to corrections of order
${v_w^2 / \Gamma_{LR} D }$ )
in equilibrium i.e. $L_1={b \over a}R_1$.
Working to leading order in ${D_L/ D_R} <<1 $
we have $R_1= -{\xi_R \over D_R}J_0$.
The right-handed lepton density is simply equal to the
effective injected flux and then putting the decay
process in equilibrium determines the density of
left-handed leptons. We can also read off $\mu_1$
from the lepton number conservation constraint
\bcsii, since each exponential solution individually
will satisfy this,
by simply putting in the equilibrium constraint.
The profile of left-handed leptons which act as a source
in the baryon number violation equation \bdotiv\ is
\eqn\profilei{\eqalign{ -{b\over a}({\xi_R \over D_R}J_0)e^{-{a+b\over a}
                                     {v_w \over D_R }z}. }}
We note that this is different from what one would infer
if one took the local densities to be determined by
the equilibrium constraint and a lepton density taken
to be carried by the right-handed leptons diffusing
in front of the wall. This calculation gives the
profile
\eqn\profileii{\eqalign{ -{b\over a+b }({\xi_R \over D_R}J_0)e^{-
                                     {v_w \over D_R }z}. }}
which, however, when integrated gives the same baryon number $B(0)$.

We can in fact generalize the simple argument used to determine
this sort of `long tail' solution when there are $n$ coupled
equations, with one species which diffuses more efficiently than
any other. Firstly we enforce the equilibrium
constraints on all the appropriate decay processes.
The equations then have  $n-A$ conservation laws of the form
of \bcsii\ and \bminusl, where $A$ is the number of decay
processes in equilibrium. We can write all but one of these
as linear combinations excluding the species which diffuses in
the long tail solution. It follows from the conservation laws
that these linear combinations can be set equal
to zero to leading order in the ratio of the diffusion constant
of any other species to the one diffusing in the long tail.
We now can determine
all amplitudes in the solution in terms of the amplitude of the
diffusing quantity. By substituting these  back in the final
conservation law involving this species we can then determine
$\mu_1$ for the solution. Finally we need to convince ourselves
that the amplitude in this exponential solution
really is equal (to leading order) to the effective injected flux
for that species. To see this one writes down the boundary condition
giving the jump in the density at the wall for  the species
which diffuses in the long tail e.g.
\eqn\profileii{\eqalign{ D_R(R_1+R_2+R_3+...)= \xi_R J_0. }}
Using the same conservation law applied to all the other solutions
individually one can see that the amplitudes of all of them ($R_2$,$R_3$ etc.)
are
next to leading order relative to the amplitude of the long tail $R_1$,
provided the amplitude of the source for all other species is
not larger than that for the diffusing species.

Let us now summarize this procedure which we will use
to extract the
information we need to calculate the baryon asymmetry
from the full set of decay equations :

$\bullet 1.$ Identify how the injected fluxes of species with diffusion
constants $D_i$ can source left-handed fermion number
significantly for the relevant wall velocity $v_w$.
Either they do so directly or through some decay process
with rate $\Gamma$ which satisfies ${D_i  \Gamma \over v_w^2} >  1$.

$\bullet 2.$ Consider only the single species with the {\it longest} diffusion
length $D_o$ which is injected (flux $J_o$)
and sources left-handed fermion number.
The amplitudes we need to determine are those in a
solution $n_i(z) = n_i(0)e^{-\mu  z}$ which describes
the diffusion of this species and
its sourcing of other species through decay processes
($\mu \sim v_w/D_0$).

$\bullet 3.$ Identify all the $A$ decay processes which satisfy
${D_o \Gamma \over v_w^2} > 1$. Impose the condition that
the appropriate signed sum of the particle density amplitudes
(assumed linearly
proportional to the chemical potentials) corresponding
to each process is equal to zero.

$\bullet 4.$ Identify the remaining $n-A$ linear combinations
of particle densities conserved by the  $A$  equilibrated processes
in step 3,
and write them so that the density of the sourcing
species appears in only one linear combination.
Set all but this latter linear combination to zero and
so determine the amplitude of all species $n_i(0)$
in terms
of the sourcing species amplitude, which is
just ${\xi_o \over D_0} J_o$ ($\xi_o$ the appropriate
persistence length).

$\bullet 5.$ To determine the root $\mu$ of the exponential
decay tail corresponding to the solution use the
remaining conserved linear combination in the
form \bminusl\ with all diffusion constants but
$D_o$ set equal to zero, and all the amplitudes
in the coefficient of $v_w$ solved in terms
of the sourcing species amplitude from step 4.

$\bullet 6.$ If the baryon number violating processes were
amongst the equilibrated processes in step 3 the final baryon asymmetry
is simply the amplitude of the baryon number in the solutions
just determined. If the baryon number violating processes
are not equilibrated, the solution for the total left-handed
fermion number must be used as a source in
$\dot B= -\Gamma_s (3B_L +L_L)$ which simply gives
$B(0)= - {\Gamma_s \over v_w \mu}(3B_L+L_L)(0)$.

For our analysis of the case of injected leptons this will
be adequate to determine the baryon number in all the different
cases of interest. For top quarks the solution of these equations
is complicated by the fact that the leading contribution from
these `long tail' solutions vanishes in various cases
so that the steps following the second above are not
applicable.
We will then have to extend our analysis to find the contributions
from other solutions which we will have to determine more carefully.
However, as we will discuss below, we will limit ourselves in this
case to a qualitative analysis of the resulting asymmetry.

{\bf 6.1 Baryon Asymmetry from an Injected Tau Lepton Flux}

We now consider in detail the baryon asymmetry produced in various
regimes by an injected flux in tau leptons. First we note
the numerical values of the parameters which we need.
In Appendix C we calculate the diffusion constants taking into account
the dominant t channel gauge boson exchange diagrams in
Figure 5

The results are
\eqn\diffcsi{\eqalign{D_L^{-1} & = 8\alpha_W^2(1+0.8\tan^4\theta_w)T \approx
  { T\over 100} \cr
  D_R^{-1} & = 28\alpha_W^2 \tan^4\theta_w T \approx {T \over 380} \cr
  D_q^{-1} & = 8 \alpha_s^2 T \approx {T \over 6}               \cr   } }
where $\theta_W$ is the Weinberg angle \ref\scattering{
We have considered only the
fermion-fermion scatterings. In ref. [31]  the
scattering rates off gluons for quarks and off $W$ bosons
for the  left-handed leptons and quarks are calculated:
these contribute roughly the
same amount to the scattering rate,
which reduces the diffusion constant by a factor of two relative
to its value in \diffcsi. There are no such scatterings for
the right-handed leptons. }.
Note that the right-handed leptons
which are coupled only through  hypercharge interactions have a very
long diffusion length:  $D_R\approx 380/T \approx 4 D_L$.

In Appendix D we calculate the rates in a plasma
for the perturbative decay processes mediated by
Higgs bosons as in Figures 3. For these we find
\eqn\decaycsi{\eqalign{\Gamma_q=0.2 \alpha_s y_t^2
\qquad \Gamma_{LR}\equiv \Gamma_{\tau_R}=2\Gamma_{\tau_L}=0.3 \alpha_w y_\tau^2
   }}
where $y_t$ and $y_{\tau}$ are the Yukawa couplings for the top
quark and tau lepton respectively. The rates for the Higgs exchange
process in Figure 4 is also calculated in Appendix D, but we
will not make use of it as it is subdominant for the cases of interest
to us.
The normalizations of these rates are defined by the convention
in \dde. Note that this means we do not incorporate the counting
factor (for color and isospin) into the rate,
just as in the case we have discussed where
$\Gamma_{LR}$ was defined so that a factor of ${1 / 2}$
appeared in front $L_L$ because of the two isospin states.
$\Gamma_A$ in \dde\ is just the collision rate per unit volume,
divided by a factor $T^3 /12$.

For the anomalous processes we have
$\nu_i$ in \dde\ differing from unity and the convention we have adopted
is an awkward one. Instead we define the rates  as the coefficients in
\eqn\sphalconvent{ \dot{B}= -  \Gamma_s(3B_L+L_L) \qquad
 \dot{\Delta}= -  \Gamma_{ss}\Delta  }
where $\Delta=B_L-B_R$. The second equation is derived by the same
arguments outlined in section 2. These give
\eqn\sphalrates{\eqalign{{\Gamma}_s= 6N_F \kappa_s \alpha_W^4 T
\approx 2\times 10^{-5}\kappa_s T  \qquad
  {\Gamma}_{ss}= 64 \kappa_{ss} \alpha_S^4 T \approx \kappa_{ss}{T\over 40}.}}
where we have adopted the standard conventions in which
$\kappa_s \alpha_W^4 T^4$
and  ${8 \over 3}\kappa_{ss} \alpha_W^4 T^4$ are  the number of
topology changing processes per unit volume per unit time, and $N_F=3$ is
the number of families.

Numerical simulations give values of $\kappa_s$ and $\kappa_{ss}$
in the range $0.1 - 1$
\ref\AmbjornAPS{J. Ambjorn, T. Askgaard, H. Porter,
M. E. Shaposhnikov, Phys. Lett. {\bf B 244}, 479 (1990);
Nucl. Phys. {\bf B 353}, 346 (1991).}

{\bf Case 1}: $v_w^2 > \Gamma_{\tau} D_R$, ${\Gamma}_s D_R$.

This is simply case 1 as treated above with $B(0)$ as in \bzeroii.
The Higgs processes do not have time to equilibrate
the injected particles with any other species, and the
baryons are produced directly from the injected left-handed
leptons and from the fraction of the injected
right-handed leptons converted to left-handed leptons.

{\bf Case 2}: $\Gamma_{\tau} D_R > v_w^2 > {\Gamma}_s D_R$.

Now the $\Gamma_{\tau}$ processes can equilibrate in the right-handed
lepton diffusion tail.
This creates a source for Higgs particles which couple in turn
to quarks. We work for definiteness
in a model in which one Higgs doublet $\phi_1$
is coupled to the charge ${2\over3}$ quarks and the second doublet
$\phi_2$ to the charge $-{1\over3}$ quarks and leptons.
We take all the processes in Figures 3 and 4 to be in equilibrium
as well as the strong sphaleron. Other permutations and
possible Higgs couplings can also be considered in a way
which will be clear from this example.
Following the general
procedure outlined above we must therefore impose
\eqn\constraintsi{\eqalign{{1\over2}\tau_L-\tau_R - {1\over 4}\phi_2 =0
\qquad & {\rm for}\quad\tau/{\rm Higgs} \cr
 {1\over6}q_L-{1\over3}b_R - {1\over 4}\phi_2 =0
\qquad & {\rm for}\quad{\rm bottom/Higgs} \cr
 {1\over6}q_L-{1\over3}t_R - {1\over 4}\phi_1 =0
\qquad & {\rm for}\quad{\rm top/Higgs} \cr
 q_L+n_L-t_R-b_R-n_R=0
\qquad & {\rm for}\quad{\rm strong}\quad {\rm sphalerons}\cr }}
where $\tau_L$ is the density of left-handed tau leptons and tau neutrinos
(minus their anti-particles), $\phi_1$ the density of Higgs
particles of both charges in doublet $\phi_1$,
$q_L$ the density of left-handed tops and bottoms, $n_L$ the density
of left-handed quarks in the lighter two generations, etc.
The numerical factors are the counting factors for color and isospin,
and also a factor of 2 for fermions relative to bosons which enters in
the conversion from chemical potential to number density. The
gauge bosons cancel out because when we subtract the equilibrium
constraints for the processes
involving all the anti-particles from those for the processes
involving the particles.
The conservation relations which we use to solve for all densities in
terms of $\tau_R$ are
\eqn\conservedi{\eqalign{q_L+t_R+b_R=0 \cr
             \tau_L+\phi_2 -b_R- {1\over4}n_L=0 \cr
                 \phi_1-t_R-{1\over4}n_L=0 \cr
                   n_L+n_R=0. \cr }}
We find these by simply determining directly the linear combinations
conserved by the processes which we have put in equilibrium.
Together \constraintsi\ and \conservedi\ give
$3B_L+L_L=q_L+n_L+\tau_L=\tau_L={182 \over 118} \tau_R$,
and following the procedure outlined above we calculate that in the
`long tail' solution the profile of $3B_L+L_L$ is
\eqn\profileiii{\eqalign{ -{182\over 118}({\xi_R \over D_R}J_0)e^{-
                                     {300\over118}{v_w \over D_R }z}. }}
which integrates to give
\eqn\bzerov{\eqalign{ B(0)={182 \over 300}{\Gamma_s D_R \over v_w^2}
                            ({\xi_R \over D_R}J_0).       }}
We note that the correction obtained over a calculation neglecting all species
except the tau leptons is negligible: the factor ${182\over 300}$
replaces ${2\over3}$, which follows from $\tau_L=2\tau_R$ and
$\mu_1=3v_wD_R^{-1}$.

{\bf Case 3}: $\Gamma_{\tau} D_R$, ${\Gamma}_s D_R >  v_w^2  >
                                                  {\Gamma}_s D_L$

In this case the sphalerons have time to reach equilibrium in the
$D_R$ tail. We must now simply add the constraint
\eqn\constraintsii{\eqalign{ q_L+n_L+\tau_L+l_L=0  }}
where $l_L$ are the left-handed leptons in the two lighter generations.
We also revise the conservations laws \conservedi\ by adding $l_L$
with the appropriate coefficient to each of them so that they also invariant
in electroweak sphaleron processes.
We then proceed as before and calculate the equilibrium density of
$B$ at the wall (which is just the amplitude of the solution) to
be
\eqn\bzerovi{\eqalign{ B(0)=-{210 \over 307}\tau_R(0)= {210\over307} ({\xi_R
\over D_R}J_0).       }}
Neglecting all species but the tau leptons we would solve
$3B_L+\tau_L=0$, $B_L=B_R$ and $\tau_L=2\tau_R$ to get
$B(0)={4\over3}( {\xi_R \over D_R}J_0)$, which differs by
a factor of two from what we have calculated including all
species.

{\bf Case 4}:  ${\Gamma}_s D_L >  v_w^2  > \Gamma_{\tau} D_R$.

Here the tau lepton-Higgs processes are too slow to allow the right-handed
particles act as a source for baryon number, but the left-handed leptons
are diffusing long enough to allow the baryon number violating processes
reach equilibrium. Since these couple the leptons to the quarks we must
impose all the constraints in \constraintsi\ above except the first one.
We use the same conservation laws as in case 3, simply removing
$\tau_L$ in the appropriate way from the second one. Solving for $B(0)$
we find
\eqn\bzerovii{\eqalign{ B(0)=-{6 \over 13}\tau_L(0)= -{6\over 13}
                                       ({\xi_L \over D_L}J_0).       }}
which again is a very minor alteration of the naive result
$B(0)=-{2\over3}\tau_L(0)$, which follows from imposing $B_L=B_R$
and $3B_L+\tau_L=0$.

{\bf 6.2 Baryon Asymmetry from an Injected Top Quark Flux}

We will not treat the case of an injected top quark flux in great
detail. Unless the walls are very thin
they are likely to be
very poorly described by our analytic calculation
 because (i) we require $L_w << m.f.p$, and
(ii) we calculated all our expressions in the
case that $m_f < m_H$, which means for the top
quark $L_w \sim m_H^{-1} < m_t^{-1} \sim 2/T$.
We will discuss these criteria in further detail
in section 8.

In the accompanying paper \JPTlongb\ we develop a new formalism
for the treatment of quarks when scattering is taken into acount.
The case of top quarks in the thin wall
regime has also been studied
by CKN in their original exposition of the `charge transport'
mechanism in \CKNctold, and our main interest here has been to
explore the interesting possibility that leptons might
source this sort of baryogenesis. For the purposes of comparison
with these previous calculations and with our calculations in
\JPTlongb\
however it is of interest to turn to this case
with the insight we have gained by studying the case of leptons.
The analysis
we  sketch can be carried through rigorously and leads to the results
which we give below.

CKN calculated the baryon asymmetry by considering the injection of a
hypercharge flux into the unbroken phase. The diffusion of
this flux was modeled numerically and the resulting profile
used as a source for a local thermal equilibrium calculation,
in which the various conserved global charges were set
to zero.
In \JPTi\ we pointed out that these constraints
are not strictly appropriate as
the local densities of global charges may float, {\it i.e.\/}
are not necessarily confined to their
globally constrained values. In the case of leptons we have
just treated this is very clear e.g. $B-L$ becomes locally
non-zero despite the fact that it is zero in the injected flux.
The method we have developed above allows us to treat this
dynamical determination of the global charges and it is
interesting to ask what effect this has in the case of quarks.
A second correction to the calculation of CKN is
that which is required because of strong sphalerons.
It was pointed out in \giud\ that CKN's local thermal
equilibrium calculation gives a null result when these
processes are put in equilibrium. The method we have developed
allows us to see the effect of a finite strong sphaleron
rate and calcualte the correction it leads to.

We will assume again that we have a model in which one doublet
is coupled to the charge ${2\over3}$ quarks and
the other to the charge $-{1\over3}$ quarks and leptons.
We will take the Yukawa couplings for the latter to be small so
that the only Higgs coupling which is relevant over the
time in which particles are diffusing is the top quark one.

The simplest case we can treat is when $v_w^2 > \Gamma  D_q$
for all decay processes. Then the injected left-handed baryon number
drives $B$ violation directly, and is caught before there is any significant
coupling to any other species.
We read off
\eqn\bzeroiix{\eqalign{ B(0)=-{3{\Gamma}_s D_q \over v_w^2}
                                       ({\xi \over D_q}J_0).       }}
where $\xi$ is as before the persistence length of the injected current.

However from \diffcsi, \decaycsi\ and \sphalrates\ we see that the
equilibrium condition $v_w^2<\Gamma D_q$ is actually satisfied for
typical non-relativistic velocities $v_w \sim 0.1$ for both the
top/Higgs processes  ($v_w^2<{y_t^2 \over 8}$)
and the strong sphalerons ($v_w^2<{\kappa_{ss} \over 9}$).
We cannot proceed however precisely as we did in
the analysis of an injected lepton flux for two reasons:

$\bullet$ If we take the
left and right-handed quarks to have the same diffusion
constant the conservation law (in the absence of
weak sphalerons)
\eqn\bcons{\eqalign{D_q B'+v_w B = 0}}
for baryon density $B$ forces $B=0$ everywhere. In the `long tail'
solution in which we impose the equilibrium constraints on
the strong sphalerons we have  $B_L=B_R$ and therefore,
as explained in section 2, since $B_L=L_L=0$,
there  is no force driving $B$ violation.

$\bullet$ Our analysis relied on the assumption
that the sourced species which diffuses furthest
is directly sourced by the injected current.
This is not true now because the Higgs particles which
are sourced indirectly by the decay processes have a
diffusion constant $D_{\phi} \sim D_L$ (for the left-handed lepton),
much larger than that of the directly sourced quarks.

To disentangle the implications of these two points we first
neglect the Higgs processes completely. The first point is
true irrespective of whether the Higgs processes are turned
on or not, and when we have clarified its implications we
will return to this second point.
We consider then the equations for the left and right-handed
baryons to which our diffusion equations reduce in the absence of
the Higgs processes:
\eqn\baryonsi{\eqalign{D_{q_L} B_L''+v_w B_L' -{\Gamma_{ss}\over 2}(B_L-B_R)
                          -3\Gamma_s B_L &= \xi_L J_0 \delta' \cr
                       D_{q_R} B_R''+v_w B_R' +{\Gamma_{ss} \over 2}(B_L-B_R)
                         &= \xi_R J_0 \delta' \cr   }}
where $D_{q_L}$ and $D_{q_R}$ are the diffusion constants for left- and
right-handed quarks respectively.
We consider these  equations in two cases:

{\bf Case 1}:  $D_{q_L}=D_{q_R}=D_q$.
We add and subtract the two equations \baryonsi\ to get two equations for
$B$ and $\Delta=B_L-B_R$
\eqn\baryonsii{\eqalign{D_q B''+v_w B'-{3\Gamma_s\over 2}( B + \Delta)
                               &=  0 \cr
                       D_q \Delta''+v_w \Delta'
              -{3\Gamma_s\over 2}( B + \Delta)-\Gamma_{ss}(\Delta)
                         &= 2\xi J_0 \delta'. \cr   }}

The $\Gamma_s$ term can be neglected in the
second equation, and the solution for
$\Delta$ obtained is  ${\xi \over D_q} J_0 \exp{-\sqrt{  \Gamma_{ss} /D_q}z}$
to leading order in ${v_w /\sqrt{ \Gamma_s D_q}}$.
This is then used as a source term in the first equation for $B$,
and the resulting baryon asymmetry at the wall is
\eqn\bzeroix{\eqalign{ B(0)=-{1 \over 2}{v_w \over \sqrt{  \Gamma_{ss} D_q}}
                         {3 {\Gamma}_s D_q \over v_w^2}
                                       ({\xi \over D_q}J_0).       }}
We see that the effect of including the strong sphalerons is to
produce a suppression by the factor
${v_w / 2\sqrt{\Gamma_{ss} D_q}}\approx { v_w /.7\sqrt\kappa_{ss}}$
of the result  in \bzeroiix.

The factor of ${1 \over 2}$ can be simply understood. It
arises when we solve the second
equation in \baryonsii. The boundary conditions impose a
jump across the wall of $\xi/ D$.
When $v_w^2 > \Gamma_{ss}D_q$ the amplitude in front of the wall
is $\approx {\xi/ D}$ and the amplitude behind suppressed.
When $v_w^2< \Gamma_{ss}D_q$ the
amplitude of the solution behind is
equal and opposite to that in front, and the roots equal, so that
exactly half the injected flux appears in front of the wall,
and half behind. The explanation for this behavior is
simply that in the low velocity limit the decays play the
role of dissipating the injected asymmetry so that the motion of the wall
becomes unimportant and the solution approaches the symmetrical one.
In the other limit the injected flux is dissipated by the wall
catching up with the diffusing particles and the only stationary
solution is reached by particles piling up
in front of the wall so that
the back diffusion current exactly cancels the injected current.

{\bf Case 2}:  $D_{q_R}-D_{q_L}= \delta D \neq 0$.

In this case the argument that there is no contribution from the long
tail breaks down because $B$ is not zero
locally in front of the wall.
To treat this case we consider
the altered version of \baryonsii, in which there is now a source
for $B$, look for the contribution from a solution
with $\Delta=0$. The equation for $B$ can be shown to be
\eqn\baryonsiii{\eqalign{\overline{D}_q B''+v_w B'-{ 3 \Gamma_s\over 2}
                                                           B
                       = - {\delta D \over \overline{D}_q }\xi J_0 \delta'  }}
where we have taken   $\xi \propto D$  for the two chiralities.
The solution for the baryon density behind the wall is
\eqn\bzerox{\eqalign{ B(0)={1 \over 2}{\delta D \over \overline{D}_q }
                         { 3 \overline{\Gamma}_s D_q \over v_w^2}
                                       ({\xi \over D_q}J_0).       }}
$\overline{D}_q\approx 1/12\alpha_s^2T$ is the
average of the two diffusion constants.

If we take $\delta D$ to arise from the difference in left and
right-handed quark diffusion because of $SU(2)$ processes like those
in Figure 5,
we use our calculations for the case of leptons to estimate
 ${\delta D \over \overline{D}_q } \approx (\alpha_W/\alpha_s)^2\approx
1/20$.
A difference in the diffusion constants also results when one incorporates
the Higgs processes in Figure 3, for which the scattering
rates differ by a factor of two for the quarks of
different handedness. A naive estimate from our calculated
rates \decaycsi\ indicates that this would give $\delta D$
of the same order.
So for low wall velocities
and/or large $\kappa_{ss}$
\bzerox\ indicates that
the dominant effect can still come from the diffusion tail.
The suppression due to the strong sphalerons in
\bzeroix\ is cut off when
one takes into account
the slightly different transport properties of  the
left- and right-handed baryons.

Finally we return to the second difference we noted between the
case of quarks and that of the leptons - that the particle species
which diffuses furthest are the Higgs particles which we have
assumed not to be directly sourced by reflection from the wall.
Including the Higgs processes should not alter significantly the
solutions we have just considered. As in the case of the leptons
these processes will simply cause a redistribution of the
injected asymmetries between the species coupled through these
processes, and this will lead simply to minor numerical corrections.
The important point is that these processes do not drive
the quantity driving baryon number to zero as the strong
sphalerons do. In neglecting the Higgs processes, however,
we did overlook the new `long-tail' solutions for the Higgs
particles which are sourced indirectly through the decay processes.
Can these particles which diffuse much further than the
quarks themselves in turn source significant baryogenesis?
By examining the boundary
conditions in the way we did in the lepton case (choosing this time to
write down a conservation law involving $\phi$ and a combination of
quarks which is sourced) one
can show that the amplitude of this solution is down by
approximately ${D_q / D_{\phi}}$,
relative to the directly injected diffusion tail.
When integrated therefore, we expect, the two solutions should give
the same order contribution.
However the further suppression due to strong sphalerons which
we have just discussed will be greater (by $\sim \sqrt{{D_q / D_\phi}}$)
for the longer tail since the baryons are diffusing longer,
and therefore we anticipate that no significant additional baryogenesis
will occur
because of the efficient diffusion of Higgs particles.
This  qualitative analysis can be carried through rigorously
just as in the lepton case by fully solving the diffusion/decay equations
with the appropriate boundary conditions.

\medskip

\centerline{\bf 7. Screening}

In the treatment of the problem we have presented we have so far
entirely neglected
the effect of  hypercharge screening. The original mechanism
of this sort proposed by CKN described the problem in terms of a hypercharge
flux generated by the reflection of top quarks off the wall.
It was pointed out by Khlebnikov
\ref\Khlebnikov{S. Yu. Khlebnikov, Phys. Lett. {\bf B300}, 376 (1993);
S. Yu. Khlebnikov, Phys. Rev. {\bf D46}, 3223 (1992).} that this treatment
overlooked the fact that any such hypercharge density would in fact be
very efficiently screened by the plasma. Since it is hypercharge that
was described as driving the production of the asymmetry it appeared
that this will cause a very significant attenuation of the effect.
This criticism applies to the mechanism as we have treated it as well
If one computes the hypercharge density profile in front
wall in any of the stationary solutions to our diffusion equations
the result is non-zero.
CKN in a later paper
\ref\CKNscreening{A. G. Cohen, D. B. Kaplan and  A. E. Nelson,
Phys. Lett. {\bf B294}, 57 (1992), Bulletin Board: hep-ph/9206214.}
argued that the effect of screening
could be accounted for by doing their constraint calculation with
injected $Y$ in a different basis of
charges `orthogonal' to $Y$. One of the  charges named $X$ which is a
linear combination of hypercharge and baryon number is identified as
the appropriate injected charge which drives the $B$ violation.
Here we will attempt to clarify this question of the role of
hypercharge screening, justifying the approximation we have made
in neglecting it and outlining how it can be incorporated
in these calculations.

Consider first a system in a constrained thermal equilibrium, with
chemical potentials $\mu_i$ for species $i$.
Suppose now we apply a local potential for hypercharge $\phi_Y(x)$
in some region. The local thermal equilibrium with the same
chemical potentials is
\eqn\equilibrium{\eqalign{ f_i(p,x) = { {1} \over
        { e^ {\beta(\epsilon_i + y_i \phi_Y(x)-\mu_i)}  \pm 1 }} \quad
 { \rm where} \quad \epsilon_i=\sqrt{p^2 +m_i^2} .}}
This makes the collision integral on the right hand side of the
Boltzmann equation zero if the original chemical potentials $\mu_i$
did. This is true because shifting the chemical potential
by an amount $y_i \phi_Y(x)$ does not alter the rates of processes
which conserve hypercharge. This distribution function
describes a solution in which the potential has the effect
of drawing in an amount of each species proportional to its hypercharge,
but this is the only change  admitted to the local density of
any species. It is this local equilibrium which CKN calculated in
\CKNscreening, fixing the potential $\phi_Y(x)$ by requiring it to screen
the hypercharge density in \equilibrium\ to zero. The same $\mu_i$
as in the unscreened case are
obtained by their prescription of working with `orthogonal'
charges. These are simply the linear combinations of charges
which are unchanged by screening i.e. charges $Q_A= \Sigma_i q_A^i n_i$
in which $\phi_Y(x)$ cancels out because
$Tr (QY) \equiv \Sigma_i k_i q_A^i y_i  =0$, where $k_i$ is
a statistical factor which is one for fermions and two for
bosons in the massless approximation. Imposing the same
constraints given by a calculated injected flux
on these charges
amounts to solving precisely the same set of linear equations
as in the unscreened case, except for the one setting
the hypercharge to zero. Quite simply one is taking the
previous solution and superimposing on it the extra screening densities
which leave rates unchanged. In particular one finds that the sphaleron
rate is unchanged and hence the resulting asymmetry.

Is this a good approximation to the real process of screening?
It describes correctly a static situation where some real
physical constraint --- an injected decaying flux or a force like that
we describe in \JPTlongb\ ---
imposes the constraints on the $\mu_i$ in some fixed region.
Then this static  equilibrium \equilibrium\ satisfies the Boltzmann
equation. However the cases we are considering are not manifestly
like this.   The region in which there are perturbations induced
(described by locally varying chemical potentials $\mu_i$)
is moving and the screening is {\it dynamical}. The screening currents
respond to the injected currents and the time it takes for
them to screen the  charges is important. In particular the static
approximation breaks down because the response of different species to
the driving hypercharge field is different because they have very
different transport properties.
To see this suppose a single species is injected by
reflection from the wall.
This injected flux carries positive hypercharge
one way and compensating hypercharge the other way, which
induces hypercharge currents in the other species
across the wall. If all
these other species are very ``stiff'' in comparison to the
original species i.e. have much smaller diffusion constants,
the screening will impede the diffusion
we have described. However, if there are species with comparable
transport properties (as there are in the electroweak plasma),
the injected flux can be screened by
these species and the effect should be simply that the
injected flux drags along screening charges with it, with
its behaviour little altered from that we have described.
This is the picture which we will now briefly support with some
simple quantitative arguments.

The way to include this dynamics is simply to incorporate screening in
our calculations. We do this by adding the number
current  induced by the hypercharge field
to the diffusion current using Ohm's Law so that we have
\eqn\scurrents{ \vec{J}= \vec{J}_d + \vec{J}_s=
               -D\vec{\nabla}n +\sigma \vec {E} }
where  $\vec{E}$ is the
hypercharge field induced by the local densities and
$\sigma$ is the conductivity. Note that this definition
of conductivity differs from the usual one by a hypercharge factor $y$,
because it relates the number current rather than the charge
current to the field.
The conductivity may be related to the diffusion constant
using the derivation
of the diffusion constant given in Appendix C
(see also \JPTi). The one difference is that the space
varying chemical potential is replaced by the hypercharge potential.
One arrives at a relation between the induced number current and
the electric field which is exactly the same except
as that between the diffusion current and the
gradient in the chemical potential, with
\eqn\conductivity{\sigma= { T^2 \over 6} y D }
Incorporating this extra term in the derivation of the
diffusion/decay equations \dde\
simply results, after using
Gauss' law, in an extra term  $-4\pi\sigma_i Y$ in
each equation.
So we have a set of equations exactly like those
we analyzed  except for this extra coupling between all
the species, which greatly
complicates the analysis we presented.

In the  limit in
which all diffusion constants are equal, a general analysis
is possible. Taking
linear combinations we can rewrite the $N$ equations
(for $N$ species)
as $N-1$ equations in which
the screening term cancels out, and one equation for the
total hypercharge $Y$ in which the decay terms cancel
out (because total hypercharge is conserved).
This corresponds precisely
to going to the `orthogonal' basis described by CKN.
In this case then the only alteration to our equations
is to add the screening term $-4\pi \Sigma_i y_i \sigma_i Y$
(where the sum is over species) to
the hypercharge propagation equation. This term is formally
just like another decay process term and we can follow
the sort of analysis given in section 6. The
``equilibrium'' condition (now for $Y=0$) becomes
${(v_w^2 / 4 \pi \Sigma_i \sigma_i D)}<<1$.
Thus for $v_w < \sqrt{{2 \pi \over 3}} \bar y DT$ (where
$\bar y^2= \Sigma_i y_i^2$) the screening will be
effective in the `long-tail' diffusion solution,
and in this regime we approach the static limit
discussed by CKN. The new set of constraints
on our `long-tail' solution will be satisfied
by superimposing a density of each species in
proportion to its hypercharge on our previous solution.
Note also that any `short-tail' solution in which
$Y \neq 0$ will be restricted to a distance
$(\sqrt{{2\pi \over 3}}\bar y T)^{-1}$ of the wall,
which is simply the inverse Debye mass in this model.

The diffusion constants of different particles in the
electroweak plasma are however, as we have seen, very different,
varying over two orders of magnitude.
The current induced in response to a hypercharge field
is therefore not proportional simply to the
hypercharge of the responding species.
The question which interests us  is which particle species
screen a given injected species and whether this can have
any very significant effects on the solutions we analysed.
In particular we would like to see the effect of the very
different transport properties of the quarks and leptons
in the screening of injected fluxes of either.
We will not carry out a general analysis of the
equations but limit ourselves to a simplified set
of equations to illustrate the intuitive argument given above.
The most important point is that the ``stiffness''
of quarks will not prevent the diffusion of leptons,
because there are other leptons around which will
screen the injected flux. Rather than drawing along
quantities of each species in proportion to their
hypercharge, the leptons will be screened by other
leptons, not by quarks.

So consider the following simple model:
\eqn\example{\eqalign { D_A A''+v_w A'
-4\pi \sigma_A(y_A A+y_B B+y_C C) &=  s_A \delta' \cr
D_B B''+v_w B'
-4 \pi \sigma_B(y_A A+y_B B+y_C C) &= 0 \cr
D_C C''+v_w C' -4 \pi \sigma_C(y_A A+y_B B+y_C C) &=  0. \cr }}
The species $A$ is sourced, while the species $B$ and $C$
screen. We have not included a decay term as this should
not be important because all such decays are hypercharge
conserving.
Let us now study the roots of these equations and
compare them to the unscreened analogue in
which $A$ simply diffuses as described by its own
diffusion constant (i.e. in the solution with
root ${v_w/D_A}$).

To obtain an equation for the
root of the screened solution
we combine \example\ to get two
equations without the
screening term which can be integrated once
to give two first order equations.
Requiring $e^{-\lambda_Dz}$ to be a solution
and imposing $Y=0$ gives
\eqn\screeningroot{\eqalign{ & D_A D_B D_C (\sum_I y_I^2)\lambda_D^2 \cr
& - v_w\bigl [ y_A^2 D_A (D_B+D_C)+ y_B^2 D_B (D_C+D_A)+
   y_C^2 D_C (D_A+D_B) \bigr ]\lambda_D\cr
& +  v_w^2 \bigl [ D_A y_A^2+ D_B y_B^2 + D_C y_C^2
 \bigr ]=0\cr
}}
Solving this equation in the limit
$D_C<< D_A, D_B$ we find a `short' root involving
$D_C$ and a `long' root
\eqn\screeningrootii{\lambda_L= v_w{D_A y_A^2+D_B y_B^2\over
D_A D_B (y_A^2+y_B^2)} \rightarrow \cases { {v_w\over D}
 & for $D_A=D_B=D$ \cr
{v_w \over D_B } {y_A^2 \over y_A^2+y_B^2 } & for $D_A >>D_B$ \cr} }
We see that if there is one species $C$ which has a very small
diffusion constant, this does not shorten the previous
solution provided there is a second species $B$ available to screen
which has comparable (or better) diffusion properties.

For baryogenesis we will typically want to evaluate
the integral under the diffusion tail $\int A$. In this case, to leading
order in $D_C/D$, species $C$ does not screen in the long tail. Using this
and the fact that the main contribution  comes from the longer tail,
one obtains
 $\int A={s_A\over v_w}{y_B^2\over y_A^2+y_B^2}$.
The suppression with respect to the unscreened case
for which $\int A={s_A\over v_w}$ is due to the screening with species $B$.
The second limit in \screeningrootii\ shows that the diffusion solution for
$A$ does not survive if both species have
much  shorter diffusion lengths --- the solution is
forced to follow the behaviour of the least `stiff'
of the screening species.

We now solve \screeningroot\ in the limit  $D_A\sim D_B<<D_C$.
We assume $D_A=D_B=D$ which considerably simplifies
potentially cumbersome algebra but still models the relevant effects
of screening.
This  case is more complicated  because there are two comparable roots
\eqn\screeningrootiii{\lambda_S=  {v_w\over D}\qquad\qquad
\lambda_L= {v_w\over D}{y_C^2\over\sum_I y_I^2}
}
A careful analysis using the appropriate boundary
conditions and imposing the condition of zero total
hypercharge with the  ansatz
$I=I_S\exp [-\lambda_S z]+I_L\exp [-\lambda_L z]$ for $z>0$ and $I=0$ for $z<0$
(where $I=A,\; B,\; C$) permits us to evaluate the amplitudes.
We find that in the shorter tail ($\lambda_S$) only species $B$
screens;  $C$ does not
screen at all (i.e. its amplitude is suppressed by $D/D_C$).
In the longer tail ($\lambda_L$)
something rather unexpected happens: species $B$
anti-screens and $C$ screens such that $y_B B/y_C C=-y_B^2/(y_A^2+y_B^2)$.
To answer how screening affects baryogenesis we again
as above study $\int A$.
Of course, if species $B$ and $C$ carry left handed
fermion number, the final
baryon asymmetry should include terms proportional to
$\int B$ and $\int C$, but let us ignore this complication.
Recall that if one species (B) screens
with $D_A=D_B=D$ we have
$\int A={s_A\over v_w}\bigl (y_B^2/ y_A^2+y_B^2\bigr )$.
When one adds  a second species (C) to screen
with $D_C>>D$
one finds that the same result
$\int A $ for the contribution of the shorter tail (in which
$C=0$). But the total result including
the longer tail contribution  is just
$\int A=s_A/v_w$ so that the unscreened result
is recovered.

In the electroweak plasma
there are many species to screen, in particular there are lighter
families which have  almost identical diffusion properties to the
third family fermions.
For example, in the case of baryogenesis sourced
by the right-handed lepton diffusion described in section 6,
we expect
that the screening will be provided by all three families of
leptons (mainly by the antiparticles of the right-handed leptons).
Still however it will be only
the $\tau_R$ leptons which are converted to
the left-handed leptons which source baryogenesis and we thus
estimate in the spirit of the simple model above
that the net result would be an attenuation
by a factor of  ${2 \over 3}$.
The second case studied above  tells us that something rather nice
occurs in the top quark mediated baryogenesis. The screening with
other quarks alone has the tendency  to decrease somewhat the baryon
asymmetry. However the presence of mobile  leptons and Higgs particles
works in the direction of recovering the unscreened result.
We conclude that the corrections due to screening of the
results we calculated in section 6 should be numerical changes
of order unity \footnote{$^\dagger$}{A recent study of this
question
\ref\ck{J. Cline and K. Kainulainen, preprint hep-ph/9506285.}
has developed the present analysis further
quantitively and reaches the same conclusion.}.

\medskip

\centerline{\bf 8. Baryon to Entropy Ratio and Validity of Calculations}

We now calculate the baryon asymmetry in
its  standard form before comparing  the
various cases we have considered. Assuming the asymmetry at the
wall to be frozen in as it passes behind the wall i.e. that
the weak sphaleron is switched off just behind the wall, we
need simply divide $B(0)$ by the entropy density
$s={2\pi^2 \over 45}g_*T^3$ ($g_*\approx 100$ the number of relativistic
degrees of freedom) to get the standard baryon number
to entropy ratio. Using the flux in \rfluxqmi\ we find, for
the most interesting parameter ranges,
\eqn\bauthin{ {n_B \over s}=
{\xi \over D}{45 \over 4g_* \pi^4} v_w ({m_f \over T})^2({m_H \over T})
\Theta_{CP} \cases {
-{\Gamma_s D_L \over v_w^2} & $ v_w^2 > \Gamma_\tau D_R,  \Gamma_s D_L$ \cr
+{2 \over 3} {\Gamma_s D_R \over v_w^2} &
$ \Gamma_\tau D_R > v_w^2 > \Gamma_s D_L$\cr
+{2 \over 3} & $ \Gamma_\tau D_R, \Gamma_s D_R > v_w^2 $ \cr
-{1\over 2}{v_w \over
\sqrt {  \Gamma_{ss} D_q} } {3 \Gamma_s D_q \over v_w^2} & $
1 > \sqrt { { v_w^2 \over  \Gamma_{ss} D_q} } > {\delta D_q \over D_q}$\cr
-{1\over 2} {\delta D_q \over D_q}{ 3\Gamma_s D_q \over v_w^2} & $
{\delta D_q \over D_q} > \sqrt{ { v_w^2 \over  \Gamma_{ss} D_q} },
v_w^2 > \Gamma_s D_q$\cr }}
where the first three cases are for leptons and the other two for quarks.
In each case $\xi$ and $D$ apply to the appropriate fermion
(left-handed lepton in first case, right-handed lepton in following
two cases, quarks in final two cases). $\Theta_{CP}$ was defined by
$\int {\cal I}m\,[m] dz={m_f \over m_H}\Theta_{CP}$.

As remarked earlier the mass $m_f$ is the tree-level finite
temperature mass,
which we should take not to be the zero temperature mass
but $y_f\phi(T)\sim y_f T$ where $y_f$
is the Yukawa coupling of the fermion.
In particular for leptons in a two doublet model this
parameter is not constrained to its standard
model value by phenomenology
\ref\yukawa{For a discussion of phenomenological
constraints see the article by I. Bigi et al. in the volume cited in \review.}.

The way we have written the result in \bauthin\  breaks it up into two
pieces --- an `injected' piece on the left of the bracket and a
`conversion' factor on the right. It is clear that the
leptons do much better in terms of conversion
because of their much greater diffusion constants
and the absence of the strong sphaleron suppression.
The extra factor of three which the quarks gain is a
color factor which really belongs on the left, because
all three colors reflect off the barrier. The factor
ended up in the conversion factor because we defined the
injected flux for quarks in terms of baryon number.

In section 3 we discussed how $Z$ could be non-zero
in either a two doublet model or in the standard model if
a $Z$ condensate is formed on the wall as described in \snnt.
In the former case it is appropriate to take $\Theta_{CP}$
to be simply the integrated phase change across a single
bubble wall as this will be determined to be the same
on every bubble wall by the effective potential,
and we can take $\Theta_{CP} \sim 1$  consistently
with phenomenology.

In the
$Z$ condensate case $\Theta_{CP}$ will contain some
 suppression (possibly many orders of magnitude)
which will depend on the details of how
one sign of the spontaneously formed condensate
comes to dominate over the other as the transition is completed.
This will be discussed in \turoknasser.

We note that for $v_w^2 < \Gamma_s D$ the asymmetry goes as
${1 / v_w}$. This is cut off if the weak anomalous
processes have time to reach equilibrium. For  velocities
$v_w < ({\Gamma_s D_{l_R}})^{1/2}
\approx 0.75 \sqrt{\kappa_s}\alpha_w\cot^2\theta_W$
({\it cf.\/}  \diffcsi\ and {\sphalrates\/}) the asymmetry goes as $v_w$.
Note that for the right-handed leptons with their very long
diffusion tail this means $v_w < 0.1 \sqrt{\kappa_s}$ so that
the third expression above can be appropriate for modest velocities.
For quarks on the other hand
$v_w < ({\Gamma_s D_q})^{1/2}
\approx 1.2 \sqrt{\kappa_s}\alpha_w^2/\alpha_s\sim \sqrt{\kappa_s}/100$
requires velocities smaller than those in the calculated
range $v_w \sim 0.1 - 1$.
Recall that $\kappa_s\sim 0.1 - 1$.

Before using \bauthin\ to calculate some
numerical values, and in order to make a detailed
comparison between the cases of
an injected quark flux and an injected lepton flux,
we return to the conditions for
the validity of \bauthin.
 In writing it down we have used the
injected flux which is calculated from
a reflection calculation which involved various assumptions.
Firstly,
we solved the Dirac equation for a free fermion by expanding it
perturbatively in $m_f L_w$. This expansion is valid provided

$\bullet$ {\it Condition 1.} $L_w<m_f^{-1}$ (perturbative expansion)

This condition  can be avoided by solving the Dirac equation exactly
for a given wall profile (Higgs mass profile) and then treating the
imaginary part  of the mass
as a perturbation (see \funakubo, \FarrarShaposhnikov).
For $L_w\gg m_f^{-1}$, as discussed in section 3, we expect a
strong WKB suppression (for typical monotonic wall
ansatzes, to which we restrict ourselves).

Secondly, we treated the fermion as free on the wall.
A typical particle in the injected asymmetry
current has a momentum $p_z\approx E v_z\approx m_H\approx 2/L_w$, with
typical energy $E\sim 2T$. The distance
the particle then advances before it scatters once, given by
$v_z\gamma_f^{-1}$,  should be larger than the wall thickness $L_w$.
This gives

$\bullet$ {\it Condition 2.} $L_w<(1/T\gamma_f)^{1/2}$  (free fermion)

One might also wonder whether a more stringent condition exists -
namely whether collisions occur on a length scale shorter than
the de Broglie wavelength of the particles. For particles
of momenta $p_z \sim m_f$, the de Broglie wavelength is substantially
larger than $m_H^{-1}$, and a correct quantum treatment must include
collisions. However, as noted above, the reflected asymmetric
flux is dominated by over-barrier reflection of particles with momenta
$p_z \sim m_H$, and for these Condition 2 suffices.

We note that this condition  is a stronger constraint than the naive
$L_w < \gamma_f^{-1}$ which we used in section 4\footnote{$^\dagger$}
{We are grateful to Larry McLerran for a discussion of this point.}.
If condition 2 is not satisfied one can estimate the
resultant phase space suppression.

Particles
with a large incident angle on the wall ($p_\perp\sim p_z$)
are nonrelativistic so we can use $p_z\approx m_f v_z\approx 2/L_w$.
The condition that these particles not scatter leads then  to a
weaker version of
condition 2: $L_w<(2/m_f\gamma_f)^{1/2}$. A simple argument based on
counting particles with large incident angle indicates that
in this case an additional suppression of order
$1/(L_w T)^2$  to \bauthin\ will result.
For the case of WKB reflection from a `bump' barrier this condition
is further relaxed because the reflection is dominated by
thermal particles with momenta $p_z \sim m_f$.

Consider now whether and how   the quarks and the
leptons meet these conditions. First consider the {\it top} quark.
Since its mass is so large
$m_t\sim T$, the condition 1 requires very thin walls $L_w<1/T$,
which is to be compared with the perturbative value $L_w\sim 20/T$.
Next condition 2 reads: $L_w< 2/T$. (The weaker version does not help
since $m_t\sim T$.) Hence unless the wall is very
thin
we expect additional strong suppressions for the {\it top} quark
which have not been included in \bauthin.

For the leptons the conditions are much more plausibly satisfied.
Take for example the
right handed lepton whose long diffusion tail  ends up in most cases
dominating baryon production. The above conditions   become:
1) $L_w<m_{l_R}^{-1}$; 2) $L_w<9/T$.
This means that for a standard model Yukawa coupling ($y_\tau=0.01$) and
a wall  $L_w\sim 10/T$ all conditions are met.
The  weaker version of condition 2 gives: $L_w<9/m_{\tau}^{1/2}$ so that
for thicker walls
we expect some phase space suppression.  Note that
we can allow $y_{\tau}$ to vary quite alot
(consistently with the phenomenology of two Higgs doublet models)
without violating the conditions.

In summary the differences between injected quark and lepton fluxes
are: (i) leptons gain in `conversion' in front of the wall
because of their better transport properties and the absence
of a strong sphaleron suppression which affects the quarks,
(ii) for very thin walls $\sim {1 / T}$ the
ratio of the injected asymmetries in quarks and leptons
is equal to the ratio of their Yukawa couplings
squared, (iii) for thicker walls $\sim {10 / T}$
there are suppressions for the quarks which are
absent for the leptons because of their
much smaller mean free path; the treatment we have
presented breaks down in this case, and new techniques for
computing quantum mechanical reflection in
the presence of scattering are required.

Two other comments on \bauthin\ should be noted.
The `persistence length' $\xi$ we estimated with a naive
argument to be $6Dv_i$, where $v_i$ is the injected velocity.
The uncertainty in $\xi$ reflects the limitation
of our calculation.
To calculate this parameter more precisely
involves going beyond the diffusion approximation to
determine exactly how the reflected asymmetry sources
the diffusion equation. This involves modelling how the
very specific momentum space distribution of the injected
flux thermalizes as it moves away from the wall. The result
is clearly ansatz dependent, as was manifest in our estimate
for the quantum mechanical and WKB cases (for which the
appropraite $v_i$ were quite different).
For a recent study of this question and a discussion of the
limitations of the diffusion approximation
see \ref\cline{J. Cline,
 Phys. Lett. {\bf B338} (1994) 263,
 hep-ph/9405365.}.
Finally we recall that our diffusion approximation
relied on taking  $v_w< v_s\sim 1/\sqrt 3$,
the speed of sound in the
plasma. For highly relativistic velocities
the particles do not have time to thermalize before they
are caught by the wall, and clearly the asymmetry will
be considerably attenuated. For $v_w>v_s$ the diffusion tail
does not exist because particles cannot diffuse faster than $v_s$.
In this case we expect local mechanisms to be the dominant sources of
baryogenesis.

There is one aspect of the free
particle approximation which we have not
discussed - the neglect of the contribution of the thermal
gauge boson excitations to the fermion
self energies \ref\Klimov{V. V. Klimov,  Yad. Fiz. {\bf 33}, 1734(1981);
Sov. J. Nucl. Phys. {\bf 33}. 934 (1981),
[Sov. Phys. JETP {\bf 55}, 199 (1982)].},
\ref\Weldon{H. A. Weldon, Phys. Rev.
{\bf D26}, 1394 and 2789 (1982).}. Weldon gave a
Dirac equation incorporating this and one can argue that this is the
relevant equation for fermionic excitations if one is interested
(as we are) in processes whose time scale is long compared to
the response time of the gauge plasma $\sim 1/gT$, but short in comparison
to the scales on which particles scatter, which is typically given by
$\sim 1/g^2T$ or $1/g^4\ln(1/g^2)T$ depending on particles' energy.
The question  of interest is then how the analysis of the
reflection problem will be modified. We present in Appendix B
a set of  manipulations of this finite temperature  Dirac
equation which lead to
a real space Dirac equation, the starting point
for an analysis in terms of relection coefficients.
We write the equation in a form in which one can see simply
how the modification of the fermion self energies at
finite temperature enters. We find that the mass matrix which relates left and
right handed particles becomes off-diagonal as a consequence of thermal
effects. These off-diagonal elements are
of order $\alpha_w (LT)^2$ and
hence small for a sufficiently thin wall. For
realistic wall thicknesses however the corrections
may be large and further analysis is
required.

Finally we turn to the numerical evaluation of \bauthin\
in a few cases, to illustrate that our final result gives
asymmetries typically compatible with the observed value.
For the standard model Yukawa couplings $m_f/T\approx y_f$ the prefactor
in \bauthin\
reads $g_*^{-1}v_w(y_f/L_w T)^2\Theta_{CP}$,
we take $\xi=6Dv_i$,
$v_i$ as given in \vinj, $m_f=y_fT$ and $2m_H^{-1}=L_w$.
In the case of the  $\tau$ lepton the first formula applies even for rather
slow walls: $v_w>2.1 y_\tau\, , \sqrt\kappa_s/20$.
In this case we find
${n_B \over s}
\approx -{2\over g_*}{y_\tau^2\over (L_w T)^2}
{\kappa_s \over v_w} \alpha_w^2\Theta_{CP}$.
If we take  $y_\tau=0.01$ (standard model value), with
$L_wT\sim 20$, $g_*=100$ and $v_w\sim 0.1$,
we get $n_B/s\sim -0.6 \times 10^{-10}\kappa_s\Theta_{CP}$,
marginally compatible with the nucleosynthesis bound $(4-7)\cdot 10^{-11}$,
provided  $CP$ violation is large $\Theta_{CP}\sim 1$.

The result in other regimes can be read off simply.
In view of our comments above one of the most interesting regimes
seems to be that of intermediate Yukawa couplings $y_\tau \sim 0.1$,
which is described by the second case above. The answer is larger by
$3 \times 10^2$ than the case of standard model Yukawa lepton coupling.
In this case we expect the reflection
calculation to be reliable, as the conditions discussed above still
hold. An extra factor 3 was obtained from the conversion of
the injected right-handed asymmetry to baryons.

\medskip

\centerline{\bf 9. Conclusion}

In this paper we have developed a new  analytic formalism to treat
the diffusion and decay of a chiral flux injected by reflection
from a CP violating bubble wall. Our formalism allows us to
fix dynamically all the injected quantum numbers rather than
constrain them locally as was done in the previous calculation
of this mechanism in the case of quarks. We have emphasized the
case of leptons, in particular the interesting role that could be
played by the right-handed leptons when the lepton Yukawa
coupling is larger than its standard model value. We have also
briefly examined the case of quark reflection, determining the
suppression which results from strong anomalous processes.

There are many interesting issues which remain outstanding
some of which we have discussed briefly
(See for example \ref\ckk{J. Cline, K. Kainulainen and A. Vischer,
hep-ph/9506284}).
The injection of
the current could be modeled in a more sophisticated way
\cline. Our calculations apply in the limit when
the wall velocity is not  relativistic, but we anticipate
they should be accurate until close to the
speed of sound in the plasma. It would be
interesting to understand the highly relativistic case
in more detail too. The `conversion' will clearly be very
much less efficient but some of this may be
compensated for by an enhancement of the reflected
asymmetry as the moving wall will be Lorentz
contracted.

\medskip

\centerline {\bf Appendix A. Quantum Mechanical Reflection
Coefficients}

We start from equation
derived in the text
\eqn\diraciii{i\partial_z \xi_{\pm} =
 \pmatrix{ \hat{E} \pm g_A Z   & m \cr
     -m & -(\hat{E} \pm g_A Z ) \cr}
    \xi_{\pm}   \qquad \qquad  S^z = \pm {1\over 2} }
from the Lagrangian \redefi, in the frame in which the wall
is at rest ($Z= Z(z)$ and $m= m(z)$) and in which the transverse
momentum $p_\perp$ has been boosted away. $\hat{E}$ is the
energy of the incident particle in this frame,
related to the energy $E$ in the unboosted
frame by $\hat{E} = \sqrt{E^2-p_\perp^2}$.
We will consider particles incident on the wall ($0>z>z_0$)
from the unbroken phase  on the right ($z >0$).
At $z>z_0$ we take the mass to have its
(real) broken phase value $m_0$.

We first perform the global rotation
\eqn\rotation{\xi_\pm \rightarrow
{\rm e}^{\pm i \sigma_3 g_A \int_{z_0}^z Z
dz}\xi_\pm }
so that the redefined $\xi_\pm$ obey
\eqn\diraciv{i\partial_z \xi_{\pm} =Q_{\pm} \xi_{\pm}
               \qquad \qquad  S^z = \pm {1\over 2} }
where
\eqn\diracv{Q_{\pm}(z)=\pmatrix{ \hat{E}    & m_\pm \cr
           -m_\pm^* & -\hat{E}  \cr} }
and $m_\pm=me^{\pm 2i g_A \int_{z_0}^z Z dz}$.
This is precisely the form in which the problem is analyzed
by CKN in \CKNctold, an analysis which we follow until
$(86)$ below.
The solution of \diraciii\ can be written
\eqn\pathord{ \xi_\pm (z)= P e^{ -i \int_{z'}^z Q_\pm (z) dz}\xi_\pm
(z') }
where P indicates path ordering.
The upper component of $\xi_\pm$ is the left-moving eigenstate in
the unbroken phase, so to describe the reflection of an incident
left-handed (L) particle we put $\xi_+(0) = \left(1 \atop
\bar{r}\right)$
and $\xi_+(z_0) = \bar{t} D^{-1} \left(1 \atop 0\right)$. $D$ is the
matrix which diagonalizes $Q_\pm=Q_o$ in the broken phase,
$DQ_oD^{-1}= \sqrt {\hat{E}^2- m_o^2}\sigma_3$.
$\bar{r}$ and $\bar{t}$ are
the reflection and tranmission amplitudes.
We will analyze only this case, extracting the reflection
coefficient for incident $\bar {L}$ (anti-particle of $L$)
and incident right-handed particle $R$ by the
substitution $m \rightarrow m^{*}$.
Using these definitions and the explicit form of the solution
\pathord\ we have
\eqn\derivi{D \Omega_o \left(1 \atop \bar{r}\right)= \left(\bar{t}
\atop 0\right)  }
where $\Omega(z)=P e^{ -i \int_0^{z} Q_+ (z) dz}$
and $\Omega_o=\Omega(z_0)=
\pmatrix { \alpha & \beta \cr \beta^* & \alpha^* \cr}$, where
$\alpha$, $\beta$ are complex numbers to be determined.
$D$ can be obtained as
$\pmatrix { c & s \cr s & c \cr}$ where $s=\sinh \theta$,
$\sinh 2\theta={m_o \over \sqrt{\hat{E}^2 - m_o^2}}$ and
$c=\cosh \theta$,
$\cosh 2\theta ={\hat{E} \over \sqrt{\hat{E}^2 - m_o^2}}$.
We then have
\eqn\derivii{\bar{r}=-{s\alpha + c \beta^* \over s \beta + c \alpha^*},
\qquad \bar{t}={1 \over s \beta + c \alpha^*}  .}
We now define
\eqn\deriviia{U(z)=\pmatrix { a & b \cr b^* & a^* \cr} =e^{ i\hat{E}
\sigma_3 z}\Omega }
for which
\eqn\usat{i\partial_z U =
\pmatrix{ 0 & m_+^*e^{2i\hat{E}z} \cr -m_+^* e^{-2i\hat{E}z} & 0 \cr
}U}
or
\eqn\absat{ i\partial_z a= m_+ e^{2i\hat{E}z}b^* \qquad
            i\partial_z b =  m_+e^{2i\hat{E}z}a^* }
\noindent
which can be integrated recursively to give
\eqn\absatrec{\eqalign{a & =  1+\int_0^z m_+(z_1) {\rm e}^{2i\hat{E}z_1}
\int_0^{z_1} m_+^*(z_2){\rm e} ^{-2i\hat{E}z_2 } +..\cr
 b & =  - i\left ( \int_0^z m_+(z_1) {\rm e}^{2i\hat{E}z_1} +
\int_0^z m_+ {\rm e}^{2i\hat{E}z_1}\int_0^{z_1} m_+^*
{\rm e}^{-2i\hat{E}z_2}\int_0^{z_2} m_+
{\rm e}^{2i\hat{E}z_3}+.. \right ) \cr  }}
Now using $\alpha=\exp{-i\hat{E}z_0}a(z_0)$ and
$\beta=\exp{-i\hat{E}z_0}b(z_0)$,
expanding the expressions \derivii\ perturbatively
in the integral $\int_0^z dz m_+ (z) \exp{2i\hat{E}z} $
we find, to leading order,
\eqn\deriviv{\eqalign{ \bar{r}= -{\rm e}^{-2i\hat{E}z_0}\left ( t
+i \int_0^{z_0} m_+ {\rm e}^{-2i\hat{E}(z-{z_0})} +i t^2 \int_0^{z_0} m^*
{\rm e}^{+2i\hat{E}(z-{z_0})} \right ) }}
where $t=\tanh \theta$ ($\tanh 2\theta={m_0 \over \hat{E}}$).
Squaring to find the reflection probability
\eqn\derivv{\eqalign{ R_{L \rightarrow R} & = |{\bar r}|^2= t^2
-it(1-t^2) \left(\int_0^{z_0} dz m_+^* {\rm e}^{2i\hat{E}(z-{z_0})}
- \int_0^{z_0} dz m_+ {\rm e}^{-2i\hat{E}(z-{z_0})}\right)\cr
& = t^2 +t(1-t^2)
\left(2\int_0^{z_0} {\cal R} {\rm e}(m_+)\sin {2i\hat{E}(z-{z_0})}
- 2 \int_0^{z_0} {\cal I}m(m_+) \cos {2i\hat{E}(z-{z_0})}\right)\cr
}}
Thus, using the replacement $m \rightarrow m^*$ for
anti-particles, we obtain  the leading term for the
difference in reflection coefficients incident with
momentum $p_z=p_z(+\infty)=\hat{E}$:
\eqn\rcoeffsi{ {\cal R}(p_z)\equiv R_{L\rightarrow R}-R_{\bar
L\rightarrow \bar R}=
           -{4 t (1-t^2)\over |m_0|} \int_{-\infty}^{\infty}
{\cal I}m [m(z) m_0^*] \cos (2 p_z z) \,dz. }
This formula should be treated with caution as at next order
in the perturbative expansion  there is a t-independent term
\eqn\nextorder{ 2\int_0^{z_o} dz_1 \int_0^{z_o} dz_2
{\cal I}m [ m_+^*(z_1) m_+(z_2)] sin 2p_z(z_1 - z_2). }
Clearly the approximation \rcoeffsi\ is only valid provided
$t$ is not so small that \rcoeffsi\ is smaller than \nextorder.
In the cases we actually apply \rcoeffsi\ this is only a marginal
approximation.

\medskip

\centerline {\bf Appendix B. Finite temperature plasma effects}

The one loop finite temperature corrected dispersion relation for fermions in
a hot gauge plasma was firstly discussed by Klimov in
Ref. \Klimov. Here we  follow
Ref. {\Weldon\/} which gives a
Dirac equation in momentum space which may be written
\eqn\appxBi{\left[(1+a)P\!\!\!\!\slash +b u\!\!\! \slash - m\right]\Psi=0}
where $P^\mu$ is the four-momentum of a particle and $u^\mu$ is the
four-velocity of the plasma. In the plasma frame $u^\mu=(1,\vec 0)$,
while in the wall frame $u^\mu=(\gamma_w,-\gamma_w \vec v_w)$, and $m$
is the tree-level mass of the particle.
For $\omega< T \,,\,\, p<T$ the coefficients $a$ and $b$ read:
\eqn\appxBii{a_{L,R}={M_{L,R}^2\over p^2}
    \left[1-{\omega\over 2p}\ln{\omega+p\over \omega-p}\right]\,,\quad
   b_{L,R}={M_{L,R}^2\over p}\left[-{\omega\over p}+
     \left({\omega^2\over p^2}-1\right){1\over 2}
     \ln{\omega+p\over \omega-p}\right]\,,}
where $M_L$ and $M_R$ are the one-loop finite temperature corrected
 left- and right-handed fermion masses, and $\omega=P^\mu u_\mu$,
$p=\left[\omega^2-P^\mu P_\mu\right]^{1/2}$  {\Weldon\/}.
Since we are mainly interested in studying thermal particles with $\omega,
p\geq T$ we have evaluated the integrals $(A4)$ -- $(A5)$ in   {\Weldon\/} and
found that the relevant change to $a$ and $b$ is:
$\ln (\omega + p)\rightarrow \ln \zeta T$, where $\zeta\simeq 1-2$.

The thermal masses for the left-handed ($qL$), right-handed {\it up} ($uR$)
and right handed  {\it down} ($dR$) quarks {\Weldon\/} are
\eqn\appxBmassesQ{\eqalign{M_{qL}^2 & =
         {2\over 3}\pi\alpha_s T^2 +{3\over 8}\pi\alpha_w T^2
                 +{1\over 72} \pi \alpha_w\tan^2\theta_W T^2 \cr
         M_{uR}^2 & =
	 {2\over 3}\pi\alpha_s T^2 + 0
+{2\over 9} \pi \alpha_w\tan^2\theta_W T^2 \cr
	 M_{dR}^2 & =
	 {2\over 3}\pi\alpha_s T^2 +0
+{1\over 18} \pi \alpha_w\tan^2\theta_W T^2 \cr
}}
while for the leptons
\eqn\appxBmassesl{\eqalign{
         M_{lL}^2 & = 0 +{3\pi\over 8}\alpha_w T^2
                 +{\pi\over 8} \alpha_w\tan^2\theta_W T^2 \cr
	 M_{lR}^2 & =
	 0 + 0 +{\pi\over 2} \alpha_w\tan^2\theta_W T^2. \cr
}}
For both quarks and leptons
\eqn\appxBmassesdiff{M_{L}^2 -M_R^2\simeq {3\pi\over 8}\alpha_w T^2
}

We incorporate the effect of a classical axial field $Z_\mu$ in
\appxBi\ by a  canonical replacement:
\eqn\appxBcanonicalreplacement{P\!\!\!\!\slash\rightarrow
P\!\!\!\!\slash - g_A Z\!\!\!\!\slash\gamma_5
}
since the $Z$ field couples axially.
We wish to answer the question of how much
the results for the reflection calculation
deviate from the free particle case studied in
the main text and in Appendix A. Our analysis will be far
from exhaustive. We will content ourselves with showing that
the corrected Dirac equation can be written with some
simple assumptions in a form which is manifestly that of
the free particle case, with corrections which are small.

We consider as in the text a one-dimensional wall lying
in the $xy$ plane. In order to study the reflection problem
we need to return to a real space Dirac equation. We
do so by replacing $P_z \rightarrow -i \partial_z$ and making the
following assumption: we assume $a$ and $b$ in \appxBi\ to be
well approximated as scalar functions of
$\omega$ by taking $p=p(\omega)$ in \appxBii. This is true in
the leading order of a derivative expansion in the background fields
i.e. in the WKB limit.
Here we are however interested in the quantum-mechanical
momenta $P_z$ typically of precisely the inverse length scale $L^{-1}$
of the variation in the background. Since
$p\sim\left[P_\perp^2+P_z^2\right]^{1/2}\rightarrow
\left[P_\perp^2-\partial_z^2\right]^{1/2}$
this  assumption should be valid
provided $p >> L^{-1}$ which will be true for the
thermal particles we wish to describe. As long as we restrict
ourselves to particles incident at small glancing angles (which
dominate phase space) our real $z-$space equation should apply.

We start by re-writing \appxBi\ (with
$Z$ field included)
for the two component right-handed $\Psi_R$ and left-handed $\Psi_L$ spinors
in the chiral representation (in which  $\Psi\sim [\Psi_R, \Psi_L ]$)
\eqn\appxBdiracRL{\eqalign{ \left [i\partial_z -g_A Z\right ]\Psi_R & =
\sigma_3 \left [ (\hat E+ c_R \hat{u}^0)-c_R\vec{\sigma}\cdot \hat{\vec u}
\right ]\Psi_R +\sigma_3 m_R(z)\Psi_L \cr
 \left [i\partial_z + g_A Z\right ]\Psi_L & =
\sigma_3 \left [- (\hat E+ c_L \hat{u}^0)-c_L\vec{\sigma}\cdot \hat{\vec u}
\right ]\Psi_L-\sigma_3 m_L(z)\Psi_R \cr
}}
where  we used $\gamma_5\Psi_{R,L}=\pm \Psi_{R,L}$.
As in the free case we are again working in the wall frame in which we have
transformed away the momentum $\vec P_\perp$  parallel to the wall by a
Lorentz transformation
\eqn\appxBboost{\hat {\vec P}_\perp
= \gamma_\perp (\vec P_\perp-\vec v_\perp E)=0
}
such that
\eqn\appxBboosti{\eqalign{\vec v_\perp & = {\vec P_\perp\over E}\,,\qquad
\gamma_\perp ={E\over \hat E}\,,\qquad \hat E={\sqrt {E^2-P_\perp^2}}
}}
and
\eqn\appxBboostii{\eqalign{\hat u^0 & =\gamma_\perp u^0\,,\qquad
\hat{\vec u}_\perp=-\gamma_\perp \vec{v}_\perp {u^0}
}}
Recall that $Z^\mu= (0,0,0,Z(z))$  and $u^\mu=(u^0,0,0,u_3)$
($u^0=\gamma_w$, $u_3=-\gamma_w v_w$) are the field
and the velocity of plasma  in the wall frame. As a  last step to
obtaining  \appxBdiracRL\ we have  divided  \appxBi\ by ($1+a_{L,R}$)
so that
\eqn\appxBthermaldefs{m_{R,L}(z)={m(z)\over 1+a_{R,L}}\,,\qquad
c_{R,L}={b_{R,L}\over 1+a_{R,L}}
}
Note that \appxBdiracRL\ is  more complicated than the corresponding free
particle one since the chirality   eigenstates
$\Psi_{R,L}= \left (1\atop 0\right )$ and
$\Psi_{R,L}= \left (0\atop 1\right )$ are no longer the eigenstates of
the Dirac operator in the massless limit. Indeed the term
\eqn\appxBthermalmixing{-c_{R,L}\sigma_3\vec{\sigma}\cdot\hat{\vec u}
}
mixes the state $\left (1\atop 0\right )$ with
$\left (0\atop 1\right )$. To find the true momentum eigenstates one
diagonalizes the the Dirac operator:
\eqn\appxBdiracoperator{\eqalign{{\cal D}_R & =\sigma_3 \left [
(\hat E+ c_R \hat{u}^0)-c_R\vec{\sigma}\cdot \hat{\vec u}\right ]\cr
{\cal D}_L & = \sigma_3 \left
[- (\hat E+ c_L \hat{u}^0)-c_L\vec{\sigma}\cdot \hat{\vec u}\right ]
}}
This induces a nontrivial transformation on $m_R$ and $m_L$ and they
become non-diagonal matrices, {\it i.e.\/}  a pure right-handed
(left-handed) travelling wave couples {\it via} plasma effects to both
components of the left-handed (right-handed) travelling wave
(in the presence of a non-zero tree level mass).

The momentum eigenvalues for the right-handed particles are
\eqn\appxBdispersion{P_z\equiv\lambda^{(\pm)}_R=
-c_R u_3\pm \left [ (\hat E+c_R\hat u^0)^2
-c_R^2 \hat u_\perp ^2 \right ]^{1\over 2}\equiv -c_R u_3\pm \lambda_R
}
and similarly for the left-handed fermions, with index $R\rightarrow L$.
It is trivial to re-write these relations in the plasma frame
(in which $u^\mu= (1,0,0,0)$)
\eqn\appxBdispersioni{P_z=\pm \left [ (E+c_R )^2- P_\perp ^2
\right ]^{1\over 2}\,,\qquad P_z=\pm \left [ (E+c_L)^2- P_\perp ^2
\right ]^{1\over 2}
}
Note that the high momentum limit  ($P>>M$) of these are
\eqn\appxBdispersionii{E^2=P^2+ 2M_R^2\,,\qquad
E^2=P^2+ 2M_L^2
}
Roughly speaking thermal particles  in plasma ($P\sim T$)
behave as free particles with
mass squared equal $2 M_{R,L}^2$ {\it plus\/} the tree level mass squared,
with the
important difference that plasma mass  does not  mix chiralities

Having found the eigenvalues \appxBdispersion\ for ${\cal D}_{R,L}$,
it is now straightforward to construct the matrix that diagonalizes them
\eqn\appxBdiagonamization{{\cal R}_R=
\left ( {\hat E+c_R \hat u^0\over \lambda_R}\right )^{1\over 2}
\left [ v^{(+)}_R,  v^{(-)}_R\right ]
}
where $ v^{(\pm )}_R$ are the eigenvectors which correspond to
$\lambda^{(\pm)}_R$
\eqn\appxBeigenvectors{\eqalign{ v^{(+)}_R & = {\cal N}_R
\pmatrix{\hat E + c_R \hat u^0 + \lambda_R\cr
c_R (\hat u^1+i \hat u^2)\cr}
\,,\qquad
v^{(-)}_R =  {\cal N}_R \pmatrix{c_R (\hat u^1-i \hat u^2)\cr
\hat E + c_R \hat u^0 + \lambda_R\cr} \cr
{\cal N}_R^{-2} & =2 (\hat E+c_R \hat u^0)
(\hat E+c_R \hat u^0+\lambda_R)
}}
and similarly for the left-handed particles.
We can now use ${\cal R}_{R,L}$ to diagonalize ${\cal D}_{L,R}$:
\eqn\appxBdiagonalization{\eqalign{ {\cal R}_R^{-1} {\cal D}_{R} {\cal R}_R
& =
{\rm diag} (-c_R\hat u^0+\lambda_R, -c_R\hat u^0-\lambda_R)\equiv d_R\cr
 {\cal R}_L^{-1} {\cal D}_{L} {\cal R}_L
& =
{\rm diag} (-c_L\hat u^0-\lambda_L, -c_L\hat u^0+\lambda_L)\equiv d_L\cr
}}
so that the left moving particles  (negative eigenvalue) correspond in this
new basis to $\left (0\atop 1\right )$ for the right-handed particles and
$\left (1\atop 0\right )$ for the left-handed particles and {\it vice versa}
for the right movers (with positive eigenvalues).

Now  \appxBdiracRL\  can be recast as
\eqn\appxBdiracRLdiag{\eqalign{ \left [i\partial_z -g_A Z\right ]
\breve\Psi_R & = d_R \breve\Psi_R + m_R(z){\cal K}_0\breve\Psi_L \cr
 \left [i\partial_z + g_A Z\right ]\breve\Psi_L & = d_L \breve\Psi_L
 -m_L(z){\cal K}_0^{-1}\breve\Psi_R \cr
}}
where we have defined ${\cal R}_R^{-1}\Psi_R=\breve\Psi_R$.
${\cal K}_0$ is an approximately diagonal  mass matrix
\eqn\appxBmatrixKo{ {\cal K}_0= {\cal R}_R^{-1}\;\sigma_3\; {\cal R}_L=
\kappa_0\sigma_3 -\eta \sigma_3\vec\sigma\cdot\hat{\vec u}_\perp
}
Indeed one can check that
\eqn\appxBmatrixKoconstants{ \kappa_0=1+o(\Delta c)^2\,,\qquad
\eta \simeq {\Delta c\over 2 (\hat E+c)} (1+o(c))
}
In order to make a comparison with the free
particle case we rotate the spinors in \appxBdiracRLdiag\
one more time as follows:
\eqn\appxBrotationofstates{\breve\Psi_R\rightarrow
\tilde\Psi_R= {\rm e}^{i\int g_A Z\, dz}\;{\rm e}^{i{d}_R z}\breve\Psi_R
\,,\qquad
\breve\Psi_L\rightarrow
\tilde\Psi_L= {\rm e}^{-i\int g_A Z\, dz}\;{\rm e}^{i{d}_L z}\breve\Psi_L
}
so that \appxBdiracRLdiag\ reduce to
\eqn\appxBdiracrotated{i\partial_z\tilde\Psi_R  =
 m_R(z){\rm e}^{2i\int g_A Z\, dz} {\cal K} \tilde\Psi_L
 \,,\qquad
 i\partial_z\tilde\Psi_L  =
- m_L(z){\rm e}^{-2i\int g_A Z\, dz} {\cal K}^{-1} \tilde\Psi_R
}
where
\eqn\appxBmatrixK{ {\cal K}= {\rm e}^{i d_R z}\;{\cal K}_0\;
{\rm e}^{-i d_L z}={\rm e}^{i\Delta c u_3 z}
\pmatrix{{\rm e}^{2i\lambda z} & -\eta {\rm e}^{2i\lambda z}
(\hat u_1-i\hat u_2) \cr
\eta {\rm e}^{-2i\lambda z} (\hat u_1+i\hat u_2) &
-{\rm e}^{-2i\lambda z}\cr}
}
Where   $\lambda ^2 =  (\hat E +c)^2 -c^2 \hat u_\perp ^2$.
The matrix ${\cal K}$ is  diagonal for free particles (when $T=0$)
and equals $\sigma_3$.
The deviation  is due to thermal effects and it is of
order $\eta\hat u_\perp\sim \alpha_w (TL)^2$.
This is small for a sufficiently thin wall, such as
we concentrate on here,  but
$O(1)$ or greater for thicker realistic walls.
We leave a fuller analysis of the effects of
these  corrections to
future  work.

\medskip

\centerline {\bf Appendix C: Diffusion Constant Calculations}

To calculate  the diffusion constant
we generalize the method outlined in \ref\Lifshitz{E. M. Lifshitz and
L. P. Pitaevskii, {\it Physical Kinetics\/}, Pergamon Press (1979).}.
 In the companion paper we present an alternative
derivation based on a local thermal equilibrium {\it
{\it ansatz}} for the distribution function, which yields a very similar
result.

We focus on the elastic scattering, $t$-channel vector
boson exchange diagrams, which are expected to dominate the
scattering process, and play the major role in limiting the
diffusion of particles (the divergence in the propagator as
$t\sim M^2$, where $M$ is the mass of the exchanged vector boson,
leads to a logarithmic enhancement of the scattering rate).

The diffusion constant $D$  measures the particle number
current induced by a gradient in the number density:
\eqn\appxAdiffusioncurrent{\vec J_d=-D \nabla n}
where $J_d$ is the diffusion (number density) current, $n$
number density. We calculate $D$ by considering a steady state
situation in which some external source and sink cause the
particle number density to vary slowly with $x$. To a first approximation,
the distribution function is just
$f_0=1/[\exp (\beta E+\xi)+1]$ (where $\beta=1/T$ is the
inverse temperature, $E=p^0$ the energy, and  $\xi$ the chemical potential),
and $\xi$ varies slowly with $x$. This distribution function  however
has no current, so we compute the deviation
$\delta f$ in the distribution function $f=f_0+\delta f$, by
solving the Boltzmann equation, and then compute the current  from
\eqn\appxAdiffusioncurrentb{\vec{J}=\int {d \!\!^{-}}^3 p \delta f\,\, {
{\vec p}\over E}
}
The perturbation $\delta f$ is of order $\tau/l$ where $\tau$
is the mean free path and $l$ the length scale over which
$\xi$ varies: the calculation is accurate in the regime where
$\tau/l <<1$.

We determine $\delta f$  from the static  Boltzmann
equation, which reduces to
\eqn\appxABoltzmannequation{d_t f=
{\vec p\over p^0}\cdot \nabla_{\vec x} f=-C(f)\, ,
}
Expanding  the collision integral $C(f)$ in terms
of $\delta f$ and expressing the $L.H.S.$ of
\appxABoltzmannequation\ in terms of $\nabla_{\vec x} n$,
we find the relation \appxAdiffusioncurrent\ from which we
read off the constant $D$.

The collision integral for two {\it in}-coming
$\{ p^\mu$, $k^\mu\}$ and two {\it out}-going particles $\{{p'}^\mu$,
${k'}^\mu\}$:
\eqn\appxAcollisionintegral{\eqalign{C(f)= &
{1 \over 2 p^0}\int_{\{p^\prime,k,k^\prime\}}
{\delta\!\!\!^{-}}^4\bigl (\Sigma p_i\bigr ) \bigl |{\cal M}\bigr |^2
{\cal P} \bigl [ f \bigr ]\, ,  \cr
 {\cal P} \bigl
[f\bigr ]= & f_p f_k (1-f_{p'}) (1-f_{k'})-f_{p'}f_{k'}(1-f_p) (1-f_k)\cr
}}
where ${\delta\!\!\!^{-}}^4 (\sum p_i )$ ensures the energy-momentum
conservation, $\int_p=\int {d\!\!\!^-}^3 p/2 p^0$,
and $ {\cal M} $ stands for  the appropriate scattering
amplitude.  For particles exchanging  $B$ vector bosons
(corresponding to the $U(1)_Y$ symmetry),
$W$ bosons  ($SU(2)_L$)  and  $g$ gluons ($SU(3)_c$), the
amplitudes are
\eqn\appxAamplitudes{\eqalign{
|{\cal M}_{B}|^2=
80 Y^2 g_1^4 { s^2 \over (t-M_B^2)^2} \cr
|{\cal M}_{W}|^2=
36 g_2^4 {   s^2 \over (t-M_W^2)^2} \cr
|{\cal M}_g|^2=32  g_3^4  {s^2+u^2 \over (t-M_g^2)^2} \cr
}}
where $Y$ is the hypercharge of the particle we are following
(in the convention $Q=T^3+Y$), $g_1$, $g_2$ and $g_3$ are the
$U(1)_Y$, $SU(2)_L$  and $SU(3)_c$ gauge couplings,
with $g_1=g_2{\rm tan} \theta_W $,
$s=(p+k)^2= 2 p\cdot k$, $t=(p-p')^2=-2 p\cdot p'$,
$u=(p-k')^2=-2 p\cdot k'$ and ${\rm sin}^2 \theta_W=0.23$, $\theta_W$ is the
Weinberg angle.  Note that in calculating \appxAamplitudes\ we have set
masses of all particles at the outer legs to zero.
This is justified since we are interested in leading order ($g^4$)
contribution to the scattering amplitudes.
The $t$-channel processes posess a logarithmic divergence when integrated
over the momenta if the exchange particle is massless.
One can cure this divergence by using the thermally corrected gluon propagator.
Since in the approximation when the scattering particles are massless and on
shell, $t\le 0$, the gluon is Debye screened.
To the leading-log accurate one can use for the propagator
$g_{\mu\nu}/(t-M_g^2)$, where $M_g(T)$ is the Debye
mass. In this approximation the dominant contribution to the scattering
integral comes from the particles with the exchange momenta of order $M_g$.
Note that so far we have not addressed the question of the difference between
the longitudinal and transverse bosons. We suspect that the contribution from
the transverse bosons, which are not Debye screened at the one-loop level,
is larger, but since we do not know what is the approprate screening mass,
we will treat all exchange bosons equally, as if they were longitudinal.
The (one-loop) Debye masses of the (longitudinal) gauge bosons are:
$M_B^2={4\pi \over 3}\alpha_w\tan^2 \theta_W T^2\approx 0.04 T^2$,
 $M_W^2={20\pi \over 3}\alpha_w T^2\approx 0.69 T^2$,
$M_g^2={8\pi}\alpha_s T^2\approx 3.6 T^2$
{\Weldon\/}, where we wrote $\alpha_w=g_2/4\pi$,
$\alpha_s=g_3/4\pi$.  Note that
the gluon thermal mass is rather large: this means that the
leading logarithm approximation calculation (which relies on
an expansion in $M/T$) may be not very accurate.

For the local equilibrium distribution functions,
${\cal P} [f_0 ]=0$ in \appxAcollisionintegral\  so that $C(f)=0$, a
consequence of the energy conservation $p^0+k^0={p'}^0+{k'}^0$.
 Now we make our key approximation ---
we assume that there is little energy transferred in a
collision, so that $p \approx p'$, which implies $k\sim k'$.
For light exchange bosons this is quite a good approximation. Nevertheless
even for exchange of  rather heavy  gluons,
this is true on average, since a particle is just as likely to
gain as to lose energy in a collision. With this, we find
to first order in $\delta f$,
${\cal P} [f]\approx (\delta f_{\vec p}-\delta f_{\vec p'}) \bigl (f^0_k
(1-f^0_k)\bigr )$, where we ignore the fluctuations in the second fluid
$\delta f_{k}$, $\delta f_{k'}$.

If locally particle concentration varies in $x$  direction, then
 the $L.H.S.$ of \appxABoltzmannequation\ becomes
${p_x \over p^0}\partial_x\xi (x)f^\prime$, where
$f^\prime=\partial f /\partial (\beta p^0+\xi)$.
This motivates the {\it Ansatz}
\eqn\appxAAnsatz{\eqalign{
\delta f_{\vec{p}} &= g(p,x) {p_x\over p^0} \cr
}}
so that
$\delta f_{\,\vec p'}= g(p')  {p'_x \over p'^0}\approx
g(p){p_x \over p^0}\cos \alpha$, where
$\cos \alpha=\vec p \cdot\vec p\,'/p p'\approx\vec p \cdot\vec p\,'/p^2$.

The Boltzmann equation {\appxABoltzmannequation},
\appxAcollisionintegral\ can be now recast as
\eqn\appxABoltzmannequationb{\eqalign{f^\prime_p \partial_x\xi
\approx & -g(p)\Gamma_t  \cr
\Gamma_t= {1 \over 2 (p^0)^3} &\int_{\{p^\prime,k,k^\prime\}}
{\delta\!\!\!^{-}}^4\bigl (\sum p_i\bigr ) \bigl |{\cal M}\bigr |^2 \bigl
(- f^\prime) p\cdot p\,'
}}
Finally using
 {\appxAdiffusioncurrent},
 {\appxAdiffusioncurrentb}, and
{\appxABoltzmannequationb}, with
$n\approx n_0-T^3\xi/12$, we find
\eqn\appxAdiffusionconstant{D={12 \over T^3}\int {d\!\!\!^{-}}^3 p
 {-f_p^\prime\over \Gamma_t}\left
({p_z\over p^0}\right )^2
}
The  integration required for  $\Gamma_t$ is simple when one notices
that the integrand of the $k'$, and $p'$ integrals is Lorentz invariant
and can therefore be performed in the center of
mass ($CM$) frame, in which $\vec p +\vec k =0= \vec p\,'+\vec k\,'$,
$p^0=k^0$, ${p'}^0={k'}^0$. We compute to leading order
in $M/T$, obtaining for example for the gluon exchange case
\eqn\appxAGammat{\Gamma_t \approx {1 \over 6 \pi}{T^3 \over p^2}
 \ln \bigl({4 pT x_2\over M^2}\bigr)
}
The diffusion constant follows immediately from
{\appxAdiffusionconstant}:
\eqn\appxAdiffusionconstantsBWg{
\eqalign{D_{B}^{-1}\approx & {100\over 7\pi} \alpha_w^2 \tan^4\Theta_W Y^2 T
\ln \biggl ({32 T^2 \over M_B^2}\biggr ) \cr
D_{W}^{-1}=&   {45 \over 7\pi} \alpha_w^2 T
\ln \biggl ({32 T^2 \over M_W^2}\biggr ) \cr
D_{G}^{-1}=& {80\over 7\pi} \alpha_s^2 T
\ln \biggl ({32 T^2 \over M_G^2}\biggr ) \cr
}}
keeping  the leading order
logarithms. (We evaluated the integrals containing  {\it logarithms}
using
$\int x^n f'_x \ln (1/x)\approx \ln(1/x_n) \int x^n f'_x$,
where $x_n\approx n$ is the value which maximizes $x^n f'_x$. This
is quite a good approximation  because $\ln(1/x)$ varies slowly where $x^n
f'_x$ is large.)
These considerations lead to the following diffusion constants for the
right- and left-handed leptons and quarks
\eqn\appxAdiffusionconstantslq{\eqalign{
D_{l_R}^{-1} = & D_B^{-1} \approx {T \over 380} \cr
D_{l_L}^{-1} = & D_W^{-1} +{ 1\over 4} D_{l_R}^{-1} \approx {T \over 100} \cr
D_{q_{L,R}}^{-1}=& Y_{q_{L,R}}^2 D_{l_R}^{-1}+\epsilon_{L,R}
D_W^{-1}+D_g^{-1}  \approx {T \over 6}
}}
where for the  coupling constants we have used
$\alpha_w\tan^2\Theta_W=1/103$,
$\alpha_w=1/30$, $\alpha_s=1/7$, and for the logarithms
involving $M_B$, $M_W$ and $M_G$, the values
6.1, 3.8, 2.2 respectively.
$\epsilon_L=1$ and $\epsilon_R=0$ take
care of the fact that the right handed quarks do not couple to the
$W$ boson. To get a feeling how good is the expansion in $(M/T)^2$ we
have also evaluated the next (mass independent) term and got a correction to
the leading logarithm: $3/2$, $3/2$, $9/4$ for the $B$, $W$ and gluon exchange
respectively, which indicates that, except perhaps for the gluon case case,
this expansion is a rather good one.

\medskip

\centerline {\bf Appendix D: Calculation of Decay Rates}

In this Appendix we outline the
calculation of the decay rates due to
fermion/Higgs and Higgs exchange processes
shown in Figures 3 and 4.

The rate per particle per unit time can be
defined in the real time formalism as the integrated
scattering amplitude squared in a thermal bath of particles:
\eqn\appxBrate{\Gamma={12\over T^3}\int _{\{ p,p',k,k'\}}
f_p \tilde f_k (1+\tilde f_{p'})(1-f_{k'}){\delta\!\!\!^-}^4(p+k-p'-k')
\bigl |{\cal M}\bigr |^2
}
Factor $12/T^3=\mu/n_0 T$ ($\mu$ is the chemical potential and $n_0$ number
density for a single species) converts the rate per unit volume and time
into the rate per unit time (see (32)).
In the above   $\int p=\int {d\!\!\!^-}^3 p/2 E_p$;
$f_p$ and $\tilde f_k$ are the
 distribution functions for the {\it in}-going fermion and
 boson; $(1+\tilde f_{p'})$ and $(1-f_{k'})$ are the spin blocking for
 the {\it  out}-going particles. We assume  thermal distributions
for particles;
if not thermal, the appropriate distribution functions should be used.

First we  focus on the  calculation of the fermion/Higgs rate.
The scattering amplitudes squared for an {\it in}-coming
left-handed, right-handed lepton and a quark are:
\eqn\appxBscatteringamplitudes{\eqalign{\bigl |{\cal M}_{l_R}\bigr |^2= &
2\bigl |{\cal M}_{l_L}\bigr |^2=
\bigl ( A_W+{1\over 4}A_B\bigr ) {- t s \over (t- m_{l_L}^2)^2}
+A_B  {- t s \over (t- m_{l_R}^2)^2}  \cr
\bigl |{\cal M}_{q}\bigr |^2= &
A_q  {- t s \over (t- m_{q}^2)^2}  \cr
}}
where the  constants are for a $B$-boson (on external leg)
$A_B=16 \pi \alpha_w \tan^2\theta_W y_{l}^2$,  for a $W$-boson
$A_W=24 \pi \alpha_w  y_{l}^2$,  and for a gluon
$A_q={128\over 3}\pi  \alpha_s  y_{q}^2$,
($\theta_W =$ Weinberg angle, $y$'s
= Yukawa couplings, $\alpha_w=1/30$, $\alpha_W \tan^2\theta_W=1/103$).
For fermion masses we take  the one-loop   thermal values:
$m_{l_R}=0.13 T$, $m_{l_L}=0.33 T$, $m_{q}=0.88 T$. To evaluate the integral
in \appxBrate\ the expression we need to integrate is
\eqn\appxBintegral{\eqalign{\Gamma= & {12\over T^3}\int _{p,k}
f_p \tilde f_k {\cal I} \cr
{\cal I} = & \int_{p',k'} {\delta\!\!\!^-}^4(p+k-p'-k')
A {- t s \over (t- m^2)^2} \cr
}}
By ignoring the spin blocking we overestimate the rate by a factor $1-4$.
(The spin blocking has a value in $[1/4, 1]$ for any momenta $\vec p$,
$\vec k$, and it is approximately {\it one } ($\sim (1-2/e^3)$) for
thermal momenta $\sim 3T$.)
${\cal I}$ can be easily evaluated in the $CM$ frame (see Appendix C)
\eqn\appxBI{{\cal I}= {A\over 8\pi}\Bigl [
\ln \bigl (1+ {2 p\cdot k\over m^2}\bigr )
-1 +{1\over {1+{2p\cdot k\over m^2}}}
\Bigr ]
}
The integrals over $p$ and $k$ we  approximate using:
\eqn\appxBintegral{\eqalign{\eta_m(n)=\int d x x^m e^{n x} \ln x
={m !\over n^{m+1}}
\left [ 1+{1\over 2}+\, .. \; +{1\over m}-{\cal C}-\ln n\right ] \cr
\int d x x^m f_x \ln {1\over x}
\approx   \ln {1\over x_{m+1}}\int d x x^m f_x\, , \cr
}}
(${\cal C}=0.577$ is the Euler constant, $x_{m+1}\approx m+1$)
and obtain  the rate per unit time:
\eqn\appxBrateB{\Gamma= {\zeta_3 A T \over 64\pi^3 }
\left [\ln\biggl ({4 x_2 T^2\over m^2}\biggr )
-2-{1\over \zeta_3}{\sum_n\eta_1(n)}\right ]
}
This estimate together with
\appxBscatteringamplitudes\ gives the following rates for right-handed,
left-handed leptons and quarks
\eqn\appxBrateC{\eqalign{\Gamma_{LR}\equiv &\Gamma_{l_R} =
2 \Gamma_{l_L} = \cr
 &  {3\zeta_3\over 8\pi^2}
\alpha_w y_{l}^2
\left [\biggl (1+{1\over 6}\tan^2\theta_W\biggr )
\ln{4 x_2 T^2\over m_{l_L}^2} +
{2\over 3} \tan^2\Theta_W \ln{4 x_2 T^2\over m_{l_R}^2}
\right ] T\approx 0.28 \alpha_w y_{l}^2 T \cr
\Gamma_{q} = &  {2\zeta_3\over 3\pi^2} \alpha_s y_{q}^2
\ln{4 x_2 T^2\over m_{q_R}^2}
 T \approx 0.19  \alpha_s y_{q}^2 T\cr
}}
In order to obtain the approximate expressions for $\Gamma$'s in
\appxBrateC\  we have
used $\tan^2\Theta_W=0.29$, and the standard one-loop formula for
thermal masses of fermions $m^2=g^2 C_R T^2/8$
($C_R$ is the Casimir constant for  representation $R$). This gives
$m_{l_R}=0.13 T$, $m_{l_L}=0.33 T$, $m_q=0.88 T$, so that the
corresponding logarithms are
$\ln (8 T^2/m_{l_R}^2)=6.3$,
$\ln (8 T^2/m_{l_L}^2)=4.3$,
$\ln (8 T^2/m_{q_R}^2)=2.3$ ($x_2\approx 2$).

The  third family rates are fastest ($\tau$ lepton
and $top$ quark)  because of their large Yukawa
couplings.

In order to evaluate the Higgs exchange diagram (Fig. 4)
 we need the scattering amplitudes:
\eqn\appxBscatteringamplitudesB{\bigl |{\cal M}\bigr |^2 =
A{t^2\over (t-m_H^2)^2}
}
where $A$ takes account of  coupling constants and counting. Some care
is required to count the diagrams for quarks and leptons.
If one assumes that {\it all} fermions couple to the same
Higgs (just like in the standard model) for the third
family one gets
\eqn\appxBcounting{\eqalign{A_{t_L}= & A_{b_L}= 12 y_{t}^2 y_{b}^2 +
2(y_{t}^2+y_{b}^2)y_{\tau}^2 +\quad {\rm other \quad families}\cr
A_{t_R}= & 12 y_{t}^2 y_{b}^2 +
4y_{t}^2 y_{\tau}^2 +\quad {\rm other \quad families}\cr
A_{b_R}= & 12 y_{t}^2 y_{b}^2 +
4y_{b}^2 y_{\tau}^2 +\quad {\rm other \quad families}\cr
A_{\tau_R}= & 2A_{\tau_L}=
12(y_{t}^2+y_{b}^2)y_{\tau}^2 +\quad {\rm other \quad families}\cr
}}
We now integrate \appxBintegral\ for the Higgs exchange process in the
$CM$ frame. The only difference  in the scattering amplitude is
$(-s)\rightarrow t$:
\eqn\appxBIB{{\cal I'}={A\over 8\pi}\left [
1-{m^2\over p\cdot k}\ln\biggl ( 1+ {2p\cdot k\over m^2} \biggr )
+{1\over{1+{2p\cdot k\over m^2}   }}
\right ]
}
We can then approximate the integrals over $\vec p$ and $\vec k$ by
expanding in $(m/T)^2$. The leading order term gives
\eqn\appxBrateHiggs{\Gamma={ A T\over 512\cdot 3 \pi}
}
Taking account of counting \appxBcounting\ we finally obtain the rates
for third family:
\eqn\appxBratesHiggs{\eqalign{\Gamma_{t_L}= & \Gamma_{b_L}=
{ T\over 128\pi}\left [
 y_{t}^2 y_{b}^2 + {1\over 6}(y_{t}^2+y_{b}^2)y_{\tau}^2
\right ]  +\quad {\rm other \quad families}\cr
\Gamma_{t_R}= & { T\over 128\pi}\left [
y_{t}^2 y_{b}^2 +
{1\over 3}y_{t}^2 y_{\tau}^2 \right ]
+\quad {\rm other \quad families}\cr
\Gamma_{b}= & { T\over 128\pi}\left [
y_{t}^2 y_{b}^2 +
{1\over 3} y_{b}^2 y_{\tau}^2 \right ]
 +\quad {\rm other \quad families}\cr
\Gamma_{\tau_R}= & 2\Gamma_{\tau_L}={ T\over 128\pi}\left [
(y_{t}^2+y_{b}^2)y_{\tau}^2\right ]
 +\quad {\rm other \quad families}\cr
}}
These rates are model dependent. Indeed if one takes
perhaps by particle physics  a more motivated case in which one Higgs
couples to the  {\it up} type of quarks and the other to the {\it down}
type and leptons, the  terms $y_t^2 y_b^2$ and $y_t^2 y_\tau^2$
 in \appxBratesHiggs\ drop, altering the rate.

\medskip

\centerline{\bf Acknowledgements}

We thank K. Kainulainen, L. McLerran and M. Shaposhnikov  for discussions.
We thank the Isaac Newton Institute, Cambridge, U.K. for hospitality
while this work was being completed.
The work of N.T. and T.P. was partially	supported by
	NSF contract PHY90-21984, and the David and Lucile Packard
Foundation. M.J. is supported by a Charlotte Elizabeth Procter Fellowship.

\listrefs

\figures
\fig{1}{Classical potential seen by a particle crossing a bubble wall.
The solid line shows the mass barrier, the dashed line the
effect of a $CP$ violating condensate field on the bubble wall.
For the case shown, the net barrier is monotonic and the $WKB$ reflection
of particles shows no $CP$ violation.
}
\fig{2}{As in Figure 1, but for
the case where the net barrier is non-monotonic and the $WKB$ reflection
of particles does show $CP$ violation.
}
\fig{3}{`Decay'  processes relevant for the propagation of
a chiral tau lepton asymmetry through the plasma.
}
\fig{4}{`Decay'  processes relevant for the propagation of
a chiral top quark  asymmetry through the plasma.
}
\fig{5}{Vector boson t-channel exchange diagrams relevant for
computation of the diffusion constants of different fermion species.
}

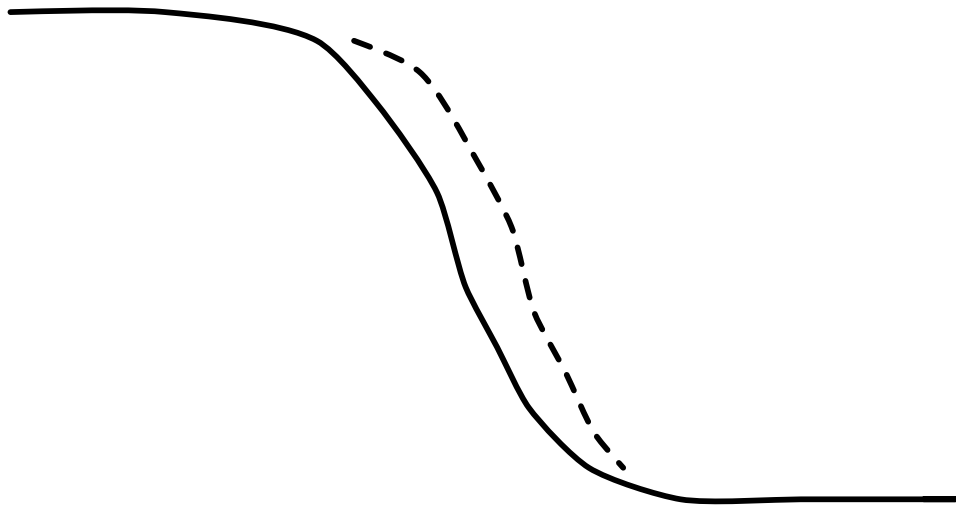
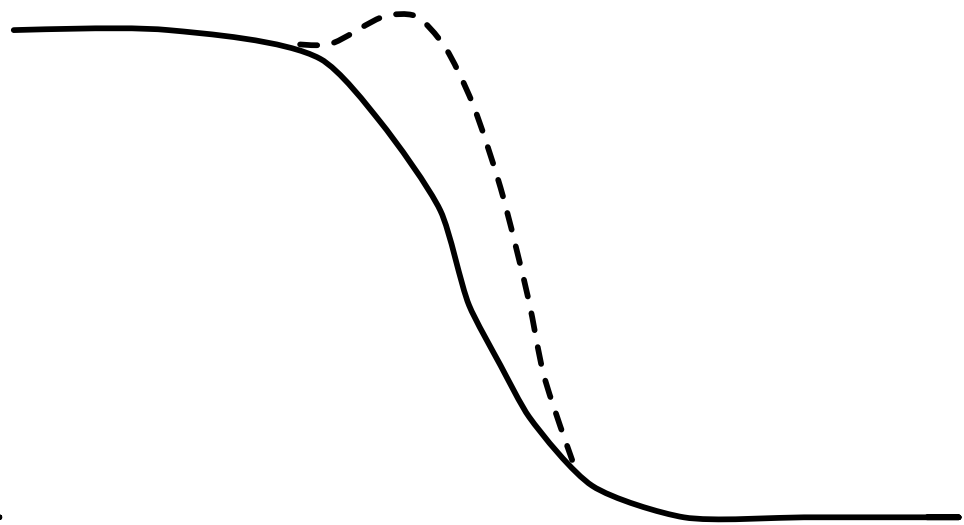
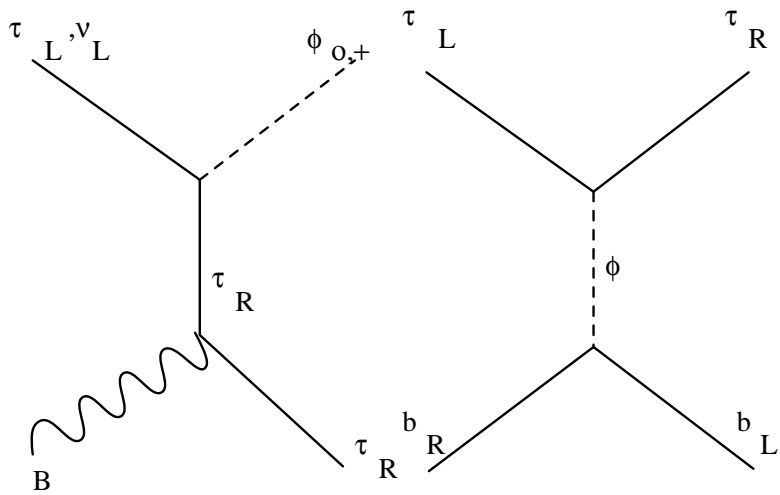
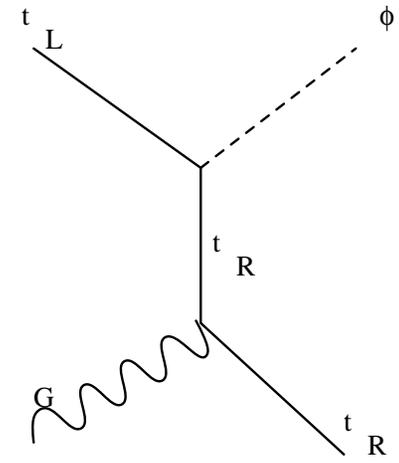
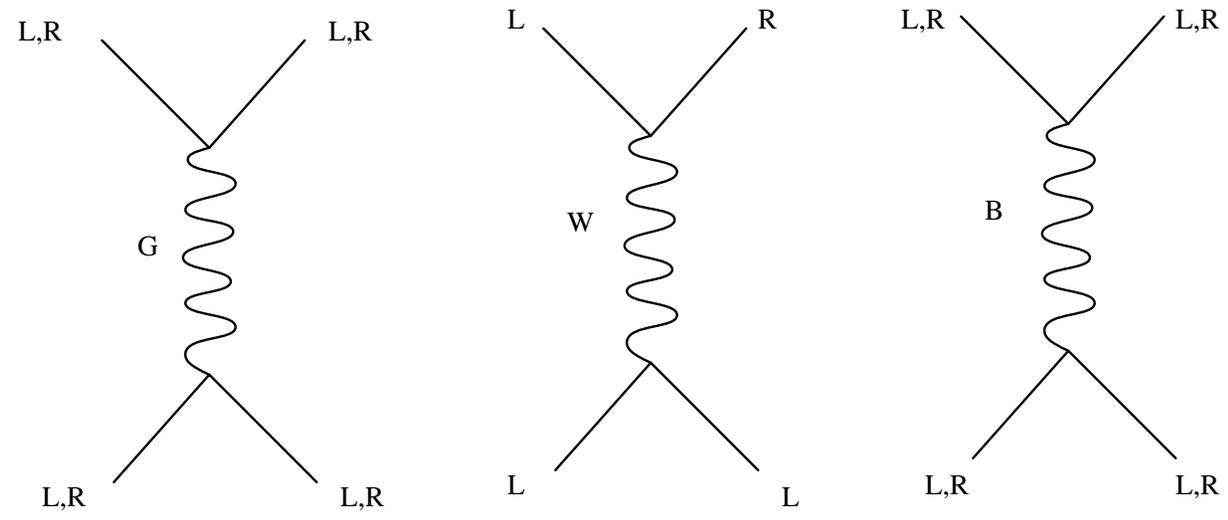
\bye